\documentclass[12pt]{article}
\usepackage{epsfig,amsfonts,amssymb}
\usepackage{hyperref}
\usepackage{cite}
\input epsf.sty
\usepackage{cite}

\newcommand{\hab}{}

\newcommand{\pii}{\pi}

\newcommand{\vq}{\xi}

\newcommand{\tree}{}

\def\ZZZ{{\hbox{ Z\kern-1.6mm Z}}}
\def\RRR{{\hbox{ R\kern-2.4mm R}}}
\def\CCC{{\hbox{ C\kern-2.0mm C}}}
\def\zzz{{\hbox{z\kern-1mm z}}}

\newcommand{\ten}{{(10)}}
\newcommand{\bet}{{( b )}}

\newcommand{\qq}{k}
\newcommand{\pp}{l}
\newcommand{\nn}{\nonumber \\}

\newcommand{\vt}{\vartheta}

\newcommand{\vtau} {\vec \tau}
\newcommand{\vj} {\vec J}
\newcommand{\vxi} {\vec \xi}
\newcommand{\vu} {\vec u}
\newcommand{\htau} {\vec \eta}
\newcommand{\vc}{\vec\chi}
\newcommand{\vpsi} {\vec \psi}

\newcommand{\qeq}{{\hbox{=\kern-2.3mm ? \kern.5mm }}}
\renewcommand{\qeq}{=}

\newcommand{\rrho}{r}
\newcommand{\bA}{{\bf A}}
\newcommand{\tx}{\wt x}
\newcommand{\bG}{{\bf G}}
\newcommand{\bF}{{\bar F}}
\newcommand{\bbb}{{\bar b}}
\newcommand{\gam}{\tau}
\newcommand{\eps}{\epsilon}
\newcommand{\vareps}{\varepsilon}
\newcommand{\ra}{\rangle}
\newcommand{\la}{\langle}
\newcommand{\T}{\chi_{T}(k)}
\newcommand{\Tm}{\chi_{T}(k')}
\newcommand{\Cn}{{\cal C}_n}
\newcommand{\vp}{\varphi}
\newcommand{\ve}{\varepsilon}
\newcommand{\tl}{\lambda}
\newcommand{\dt}{(\vec \nabla T)^2}
\newcommand{\hp}{{\wh\Phi}}
\newcommand{\hq}{{\wh Q_B}}
\newcommand{\he}{{\wh\eta_0}}
\newcommand{\ha}{{\wh{A}}}
\newcommand{\lllb}{\Bigl\langle\Bigl\langle}
\newcommand{\rrrb}{\Bigr\rangle\Bigr\rangle}
\newcommand{\tf}{\wt f}
\newcommand{\sss}{{\cal L}_{av}}
\newcommand{\bx}{\bar x}
\newcommand{\bw}{\bar w}
\newcommand{\ws}{{\wt\sigma}}
\newcommand{\wrh}{{\wt\rho}}
\newcommand{\wv}{{\wt v}}

\newcommand{\vv} {\bar v}
\newcommand{\uu} {\bar u}

\newcommand{\K}{{\rm K_1}}
\newcommand{\Kt}{{\rm \widetilde K_1}}

\newcommand{\B}{b'}
\newcommand{\C}{c\,'}
\newcommand{\bB}{\bar b'}
\newcommand{\Bu}{B_{\vec u}}
\newcommand{\VV}{{\cal V}}
\newcommand{\BB}{{\cal B}}
\newcommand{\DD}{{\cal D}}
\newcommand{\BBB}{{\cal B}}
\newcommand{\II}{{\cal I}}
\newcommand{\AAA}{{\cal A}}
\newcommand{\GG}{{\cal G}}
\newcommand{\KK}{{\cal K}}
\newcommand{\fff}{{\bf f}}
\newcommand{\ccc}{{\bf c}}
\newcommand{\FF}{{\cal F}}
\newcommand{\JJ}{{\cal J}}
\newcommand{\HH}{{\cal H}}
\newcommand{\MM}{{\cal M}}
\newcommand{\CC}{{\cal C}}
\newcommand{\bC}{{\bf C}}
\newcommand{\OO}{{\cal O}}
\newcommand{\QQ}{{\cal Q}}
\newcommand{\PP}{{\cal P}}
\newcommand{\EE}{{\cal E}}
\newcommand{\LL}{{\cal L}}

\newcommand{\XX}{{\cal X}}

 \newcommand{\rrr}{\rangle\rangle}
\newcommand{\half}{{1\over 2}}
\newcommand{\wt}{\widetilde}
\newcommand{\wh}{\widehat}
\newcommand{\wc}{\wt}
\newcommand{\wb}{\bar}
\newcommand{\RR}{{\cal R}}
\newcommand{\NN}{{\cal N}}
\newcommand{\TT}{{\cal T}}
\newcommand{\bg}{\bar g}
\newcommand{\ba}{\bar a}
\newcommand{\bc}{\bar c}
\newcommand{\bd}{\bar d}
\newcommand{\bb}{\bar b}
\newcommand{\bT}{\bar \Theta}
\newcommand{\SSS}{{\cal S}}
\newcommand{\tlx}{\left(\tilde \lambda ; X^0(0) \right)}
\newcommand{\al}{\alpha}

\newcommand{\tk}{\tilde \kappa}

\newcommand{\ppp}{\prime\prime}

\newcommand{\omk}{\omega_n(\vec k)}
\newcommand{\onk}{\omega^{(N)}_{\vec k_\perp}}
\newcommand{\tI}{\wt\II}
\newcommand{\hI}{\wh\II}
\newcommand{\nI}{\II}

\newcommand{\cp}{\check\Phi}
\newcommand{\cps}{\Psi}
\newcommand{\crh}{\check\rho}
\newcommand{\cs}{\check\sigma}
\newcommand{\cv}{\check v}
\newcommand{\com}{\check\Omega}

\newcommand{\be}{\begin{equation}}
\newcommand{\ee}{\end{equation}}
\newcommand{\ben}{\begin{eqnarray}\displaystyle}
\newcommand{\een}{\end{eqnarray}}

\newcommand{\refb}[1]{(\ref{#1})}
\newcommand{\p}{\partial}
\newcommand{\sectiono}[1]{\section{#1}\setcounter{equation}{0}}
\newcommand{\subsectiono}[1]{\subsection{#1}\setcounter{equation}{0}}

\newcommand{\zet}{\zeta}

\newcommand{\gsim}{\stackrel{>}{\sim}}
\newcommand{\lsim}{\stackrel{<}{\sim}}

\newcommand{\Lamb}{\Lambda}

\def\one{{\hbox{ 1\kern-.8mm l}}}
\def\zero{{\hbox{ 0\kern-1.5mm 0}}}

\def\wa{{\wh a}}
\def\wb{{\wh b}}
\def\wc{{\wh c}}
\def\wc{\check}
\def\wdd{{\wh d}}

\newcommand{\bi}{{\bf i}}

\renewcommand{\theequation}{\thesection.\arabic{equation}}

\newcommand{\bea}[1]{\begin{eqnarray}\label{#1} }
\newcommand{\eea}{\end{eqnarray}}

\newcommand{\wJ}{\wt J}
\newcommand{\bN}{{\bf N}}

\newcommand{\aaa}{b}

\newcommand{\eqref}{\refb}

\newcommand{\un}{{\rm u}}

\newcommand{\dotalpha}{{\dot{\alpha}}}

\newcommand{\dotbeta}{{\dot{\beta}}}

\newcommand{\dotgamma}{{\dot{\gamma}}}

\newcommand{\dalpha}{\beta}

\newcommand{\Vm}{V}

\newcommand{\gb}{G}

\newcommand{\q}{e}

\newcommand{\PPP}{{\cal P}}

\newcommand{\gold}{\VV_{\rm G}}

\newcommand{\goldc}{\VV^c_{\rm G}}

\usepackage{bm}
\usepackage[table]{xcolor}
\def\rpnote#1{{\color{magenta} #1}}
\def\arnote#1{{\color{blue} #1}}
\def\asnote#1{{\color{red} #1}}

\newcommand{\scalar}{\VV_{\rm S}} 

\newcommand{\wscalar}{\wt\VV_{\rm B}}

\newcommand{\fermion}{\VV_{\rm F}} 

\newcommand{\wfermion}{\wt\VV_{\rm F}}  

\newcommand{\wts}{\wt\Sigma}

\newcommand{\wtsp}{\wt\Sigma^c}

\newcommand{\four}{(4)}

\newcommand{\cL} {\{\hskip -4pt\{}
\newcommand{\cR} {\}\hskip -4pt\}}
\newcommand{\sL} {[\hskip -1.5pt[}
\newcommand{\sR} {]\hskip -1.5pt]}
\newcommand{\oR}{{\overline{\RR}}}

\def\figsofttwotree{

\def\JPicScale{0.6}
\ifx\JPicScale\undefined\def\JPicScale{1}\fi
\unitlength \JPicScale mm
\begin{picture}(160,90)(0,0)

\linethickness{0.3mm}
\put(109.89,44.75){\line(0,1){0.5}}
\multiput(109.88,45.75)(0.02,-0.5){1}{\line(0,-1){0.5}}
\multiput(109.84,46.24)(0.03,-0.5){1}{\line(0,-1){0.5}}
\multiput(109.8,46.74)(0.05,-0.5){1}{\line(0,-1){0.5}}
\multiput(109.73,47.23)(0.07,-0.49){1}{\line(0,-1){0.49}}
\multiput(109.65,47.72)(0.08,-0.49){1}{\line(0,-1){0.49}}
\multiput(109.55,48.21)(0.1,-0.49){1}{\line(0,-1){0.49}}
\multiput(109.43,48.7)(0.12,-0.48){1}{\line(0,-1){0.48}}
\multiput(109.3,49.18)(0.13,-0.48){1}{\line(0,-1){0.48}}
\multiput(109.15,49.65)(0.15,-0.48){1}{\line(0,-1){0.48}}
\multiput(108.99,50.12)(0.16,-0.47){1}{\line(0,-1){0.47}}
\multiput(108.81,50.59)(0.18,-0.46){1}{\line(0,-1){0.46}}
\multiput(108.62,51.04)(0.1,-0.23){2}{\line(0,-1){0.23}}
\multiput(108.41,51.5)(0.1,-0.23){2}{\line(0,-1){0.23}}
\multiput(108.18,51.94)(0.11,-0.22){2}{\line(0,-1){0.22}}
\multiput(107.94,52.38)(0.12,-0.22){2}{\line(0,-1){0.22}}
\multiput(107.69,52.8)(0.13,-0.21){2}{\line(0,-1){0.21}}
\multiput(107.42,53.22)(0.13,-0.21){2}{\line(0,-1){0.21}}
\multiput(107.14,53.63)(0.14,-0.21){2}{\line(0,-1){0.21}}
\multiput(106.84,54.04)(0.15,-0.2){2}{\line(0,-1){0.2}}
\multiput(106.54,54.43)(0.1,-0.13){3}{\line(0,-1){0.13}}
\multiput(106.21,54.81)(0.11,-0.13){3}{\line(0,-1){0.13}}
\multiput(105.88,55.18)(0.11,-0.12){3}{\line(0,-1){0.12}}
\multiput(105.53,55.53)(0.12,-0.12){3}{\line(0,-1){0.12}}
\multiput(105.18,55.88)(0.12,-0.12){3}{\line(1,0){0.12}}
\multiput(104.81,56.21)(0.12,-0.11){3}{\line(1,0){0.12}}
\multiput(104.43,56.54)(0.13,-0.11){3}{\line(1,0){0.13}}
\multiput(104.04,56.84)(0.13,-0.1){3}{\line(1,0){0.13}}
\multiput(103.63,57.14)(0.2,-0.15){2}{\line(1,0){0.2}}
\multiput(103.22,57.42)(0.21,-0.14){2}{\line(1,0){0.21}}
\multiput(102.8,57.69)(0.21,-0.13){2}{\line(1,0){0.21}}
\multiput(102.38,57.94)(0.21,-0.13){2}{\line(1,0){0.21}}
\multiput(101.94,58.18)(0.22,-0.12){2}{\line(1,0){0.22}}
\multiput(101.5,58.41)(0.22,-0.11){2}{\line(1,0){0.22}}
\multiput(101.04,58.62)(0.23,-0.1){2}{\line(1,0){0.23}}
\multiput(100.59,58.81)(0.23,-0.1){2}{\line(1,0){0.23}}
\multiput(100.12,58.99)(0.46,-0.18){1}{\line(1,0){0.46}}
\multiput(99.65,59.15)(0.47,-0.16){1}{\line(1,0){0.47}}
\multiput(99.18,59.3)(0.48,-0.15){1}{\line(1,0){0.48}}
\multiput(98.7,59.43)(0.48,-0.13){1}{\line(1,0){0.48}}
\multiput(98.21,59.55)(0.48,-0.12){1}{\line(1,0){0.48}}
\multiput(97.72,59.65)(0.49,-0.1){1}{\line(1,0){0.49}}
\multiput(97.23,59.73)(0.49,-0.08){1}{\line(1,0){0.49}}
\multiput(96.74,59.8)(0.49,-0.07){1}{\line(1,0){0.49}}
\multiput(96.24,59.84)(0.5,-0.05){1}{\line(1,0){0.5}}
\multiput(95.75,59.88)(0.5,-0.03){1}{\line(1,0){0.5}}
\multiput(95.25,59.89)(0.5,-0.02){1}{\line(1,0){0.5}}
\put(94.75,59.89){\line(1,0){0.5}}
\multiput(94.25,59.88)(0.5,0.02){1}{\line(1,0){0.5}}
\multiput(93.76,59.84)(0.5,0.03){1}{\line(1,0){0.5}}
\multiput(93.26,59.8)(0.5,0.05){1}{\line(1,0){0.5}}
\multiput(92.77,59.73)(0.49,0.07){1}{\line(1,0){0.49}}
\multiput(92.28,59.65)(0.49,0.08){1}{\line(1,0){0.49}}
\multiput(91.79,59.55)(0.49,0.1){1}{\line(1,0){0.49}}
\multiput(91.3,59.43)(0.48,0.12){1}{\line(1,0){0.48}}
\multiput(90.82,59.3)(0.48,0.13){1}{\line(1,0){0.48}}
\multiput(90.35,59.15)(0.48,0.15){1}{\line(1,0){0.48}}
\multiput(89.88,58.99)(0.47,0.16){1}{\line(1,0){0.47}}
\multiput(89.41,58.81)(0.46,0.18){1}{\line(1,0){0.46}}
\multiput(88.96,58.62)(0.23,0.1){2}{\line(1,0){0.23}}
\multiput(88.5,58.41)(0.23,0.1){2}{\line(1,0){0.23}}
\multiput(88.06,58.18)(0.22,0.11){2}{\line(1,0){0.22}}
\multiput(87.62,57.94)(0.22,0.12){2}{\line(1,0){0.22}}
\multiput(87.2,57.69)(0.21,0.13){2}{\line(1,0){0.21}}
\multiput(86.78,57.42)(0.21,0.13){2}{\line(1,0){0.21}}
\multiput(86.37,57.14)(0.21,0.14){2}{\line(1,0){0.21}}
\multiput(85.96,56.84)(0.2,0.15){2}{\line(1,0){0.2}}
\multiput(85.57,56.54)(0.13,0.1){3}{\line(1,0){0.13}}
\multiput(85.19,56.21)(0.13,0.11){3}{\line(1,0){0.13}}
\multiput(84.82,55.88)(0.12,0.11){3}{\line(1,0){0.12}}
\multiput(84.47,55.53)(0.12,0.12){3}{\line(1,0){0.12}}
\multiput(84.12,55.18)(0.12,0.12){3}{\line(0,1){0.12}}
\multiput(83.79,54.81)(0.11,0.12){3}{\line(0,1){0.12}}
\multiput(83.46,54.43)(0.11,0.13){3}{\line(0,1){0.13}}
\multiput(83.16,54.04)(0.1,0.13){3}{\line(0,1){0.13}}
\multiput(82.86,53.63)(0.15,0.2){2}{\line(0,1){0.2}}
\multiput(82.58,53.22)(0.14,0.21){2}{\line(0,1){0.21}}
\multiput(82.31,52.8)(0.13,0.21){2}{\line(0,1){0.21}}
\multiput(82.06,52.38)(0.13,0.21){2}{\line(0,1){0.21}}
\multiput(81.82,51.94)(0.12,0.22){2}{\line(0,1){0.22}}
\multiput(81.59,51.5)(0.11,0.22){2}{\line(0,1){0.22}}
\multiput(81.38,51.04)(0.1,0.23){2}{\line(0,1){0.23}}
\multiput(81.19,50.59)(0.1,0.23){2}{\line(0,1){0.23}}
\multiput(81.01,50.12)(0.18,0.46){1}{\line(0,1){0.46}}
\multiput(80.85,49.65)(0.16,0.47){1}{\line(0,1){0.47}}
\multiput(80.7,49.18)(0.15,0.48){1}{\line(0,1){0.48}}
\multiput(80.57,48.7)(0.13,0.48){1}{\line(0,1){0.48}}
\multiput(80.45,48.21)(0.12,0.48){1}{\line(0,1){0.48}}
\multiput(80.35,47.72)(0.1,0.49){1}{\line(0,1){0.49}}
\multiput(80.27,47.23)(0.08,0.49){1}{\line(0,1){0.49}}
\multiput(80.2,46.74)(0.07,0.49){1}{\line(0,1){0.49}}
\multiput(80.16,46.24)(0.05,0.5){1}{\line(0,1){0.5}}
\multiput(80.12,45.75)(0.03,0.5){1}{\line(0,1){0.5}}
\multiput(80.11,45.25)(0.02,0.5){1}{\line(0,1){0.5}}
\put(80.11,44.75){\line(0,1){0.5}}
\multiput(80.11,44.75)(0.02,-0.5){1}{\line(0,-1){0.5}}
\multiput(80.12,44.25)(0.03,-0.5){1}{\line(0,-1){0.5}}
\multiput(80.16,43.76)(0.05,-0.5){1}{\line(0,-1){0.5}}
\multiput(80.2,43.26)(0.07,-0.49){1}{\line(0,-1){0.49}}
\multiput(80.27,42.77)(0.08,-0.49){1}{\line(0,-1){0.49}}
\multiput(80.35,42.28)(0.1,-0.49){1}{\line(0,-1){0.49}}
\multiput(80.45,41.79)(0.12,-0.48){1}{\line(0,-1){0.48}}
\multiput(80.57,41.3)(0.13,-0.48){1}{\line(0,-1){0.48}}
\multiput(80.7,40.82)(0.15,-0.48){1}{\line(0,-1){0.48}}
\multiput(80.85,40.35)(0.16,-0.47){1}{\line(0,-1){0.47}}
\multiput(81.01,39.88)(0.18,-0.46){1}{\line(0,-1){0.46}}
\multiput(81.19,39.41)(0.1,-0.23){2}{\line(0,-1){0.23}}
\multiput(81.38,38.96)(0.1,-0.23){2}{\line(0,-1){0.23}}
\multiput(81.59,38.5)(0.11,-0.22){2}{\line(0,-1){0.22}}
\multiput(81.82,38.06)(0.12,-0.22){2}{\line(0,-1){0.22}}
\multiput(82.06,37.62)(0.13,-0.21){2}{\line(0,-1){0.21}}
\multiput(82.31,37.2)(0.13,-0.21){2}{\line(0,-1){0.21}}
\multiput(82.58,36.78)(0.14,-0.21){2}{\line(0,-1){0.21}}
\multiput(82.86,36.37)(0.15,-0.2){2}{\line(0,-1){0.2}}
\multiput(83.16,35.96)(0.1,-0.13){3}{\line(0,-1){0.13}}
\multiput(83.46,35.57)(0.11,-0.13){3}{\line(0,-1){0.13}}
\multiput(83.79,35.19)(0.11,-0.12){3}{\line(0,-1){0.12}}
\multiput(84.12,34.82)(0.12,-0.12){3}{\line(0,-1){0.12}}
\multiput(84.47,34.47)(0.12,-0.12){3}{\line(1,0){0.12}}
\multiput(84.82,34.12)(0.12,-0.11){3}{\line(1,0){0.12}}
\multiput(85.19,33.79)(0.13,-0.11){3}{\line(1,0){0.13}}
\multiput(85.57,33.46)(0.13,-0.1){3}{\line(1,0){0.13}}
\multiput(85.96,33.16)(0.2,-0.15){2}{\line(1,0){0.2}}
\multiput(86.37,32.86)(0.21,-0.14){2}{\line(1,0){0.21}}
\multiput(86.78,32.58)(0.21,-0.13){2}{\line(1,0){0.21}}
\multiput(87.2,32.31)(0.21,-0.13){2}{\line(1,0){0.21}}
\multiput(87.62,32.06)(0.22,-0.12){2}{\line(1,0){0.22}}
\multiput(88.06,31.82)(0.22,-0.11){2}{\line(1,0){0.22}}
\multiput(88.5,31.59)(0.23,-0.1){2}{\line(1,0){0.23}}
\multiput(88.96,31.38)(0.23,-0.1){2}{\line(1,0){0.23}}
\multiput(89.41,31.19)(0.46,-0.18){1}{\line(1,0){0.46}}
\multiput(89.88,31.01)(0.47,-0.16){1}{\line(1,0){0.47}}
\multiput(90.35,30.85)(0.48,-0.15){1}{\line(1,0){0.48}}
\multiput(90.82,30.7)(0.48,-0.13){1}{\line(1,0){0.48}}
\multiput(91.3,30.57)(0.48,-0.12){1}{\line(1,0){0.48}}
\multiput(91.79,30.45)(0.49,-0.1){1}{\line(1,0){0.49}}
\multiput(92.28,30.35)(0.49,-0.08){1}{\line(1,0){0.49}}
\multiput(92.77,30.27)(0.49,-0.07){1}{\line(1,0){0.49}}
\multiput(93.26,30.2)(0.5,-0.05){1}{\line(1,0){0.5}}
\multiput(93.76,30.16)(0.5,-0.03){1}{\line(1,0){0.5}}
\multiput(94.25,30.12)(0.5,-0.02){1}{\line(1,0){0.5}}
\put(94.75,30.11){\line(1,0){0.5}}
\multiput(95.25,30.11)(0.5,0.02){1}{\line(1,0){0.5}}
\multiput(95.75,30.12)(0.5,0.03){1}{\line(1,0){0.5}}
\multiput(96.24,30.16)(0.5,0.05){1}{\line(1,0){0.5}}
\multiput(96.74,30.2)(0.49,0.07){1}{\line(1,0){0.49}}
\multiput(97.23,30.27)(0.49,0.08){1}{\line(1,0){0.49}}
\multiput(97.72,30.35)(0.49,0.1){1}{\line(1,0){0.49}}
\multiput(98.21,30.45)(0.48,0.12){1}{\line(1,0){0.48}}
\multiput(98.7,30.57)(0.48,0.13){1}{\line(1,0){0.48}}
\multiput(99.18,30.7)(0.48,0.15){1}{\line(1,0){0.48}}
\multiput(99.65,30.85)(0.47,0.16){1}{\line(1,0){0.47}}
\multiput(100.12,31.01)(0.46,0.18){1}{\line(1,0){0.46}}
\multiput(100.59,31.19)(0.23,0.1){2}{\line(1,0){0.23}}
\multiput(101.04,31.38)(0.23,0.1){2}{\line(1,0){0.23}}
\multiput(101.5,31.59)(0.22,0.11){2}{\line(1,0){0.22}}
\multiput(101.94,31.82)(0.22,0.12){2}{\line(1,0){0.22}}
\multiput(102.38,32.06)(0.21,0.13){2}{\line(1,0){0.21}}
\multiput(102.8,32.31)(0.21,0.13){2}{\line(1,0){0.21}}
\multiput(103.22,32.58)(0.21,0.14){2}{\line(1,0){0.21}}
\multiput(103.63,32.86)(0.2,0.15){2}{\line(1,0){0.2}}
\multiput(104.04,33.16)(0.13,0.1){3}{\line(1,0){0.13}}
\multiput(104.43,33.46)(0.13,0.11){3}{\line(1,0){0.13}}
\multiput(104.81,33.79)(0.12,0.11){3}{\line(1,0){0.12}}
\multiput(105.18,34.12)(0.12,0.12){3}{\line(1,0){0.12}}
\multiput(105.53,34.47)(0.12,0.12){3}{\line(0,1){0.12}}
\multiput(105.88,34.82)(0.11,0.12){3}{\line(0,1){0.12}}
\multiput(106.21,35.19)(0.11,0.13){3}{\line(0,1){0.13}}
\multiput(106.54,35.57)(0.1,0.13){3}{\line(0,1){0.13}}
\multiput(106.84,35.96)(0.15,0.2){2}{\line(0,1){0.2}}
\multiput(107.14,36.37)(0.14,0.21){2}{\line(0,1){0.21}}
\multiput(107.42,36.78)(0.13,0.21){2}{\line(0,1){0.21}}
\multiput(107.69,37.2)(0.13,0.21){2}{\line(0,1){0.21}}
\multiput(107.94,37.62)(0.12,0.22){2}{\line(0,1){0.22}}
\multiput(108.18,38.06)(0.11,0.22){2}{\line(0,1){0.22}}
\multiput(108.41,38.5)(0.1,0.23){2}{\line(0,1){0.23}}
\multiput(108.62,38.96)(0.1,0.23){2}{\line(0,1){0.23}}
\multiput(108.81,39.41)(0.18,0.46){1}{\line(0,1){0.46}}
\multiput(108.99,39.88)(0.16,0.47){1}{\line(0,1){0.47}}
\multiput(109.15,40.35)(0.15,0.48){1}{\line(0,1){0.48}}
\multiput(109.3,40.82)(0.13,0.48){1}{\line(0,1){0.48}}
\multiput(109.43,41.3)(0.12,0.48){1}{\line(0,1){0.48}}
\multiput(109.55,41.79)(0.1,0.49){1}{\line(0,1){0.49}}
\multiput(109.65,42.28)(0.08,0.49){1}{\line(0,1){0.49}}
\multiput(109.73,42.77)(0.07,0.49){1}{\line(0,1){0.49}}
\multiput(109.8,43.26)(0.05,0.5){1}{\line(0,1){0.5}}
\multiput(109.84,43.76)(0.03,0.5){1}{\line(0,1){0.5}}
\multiput(109.88,44.25)(0.02,0.5){1}{\line(0,1){0.5}}

\linethickness{0.8mm}
\put(10,45){\line(1,0){48}}
\linethickness{0.8mm}
\put(58,45){\line(1,0){22}}
\linethickness{0.8mm}
\multiput(109,50)(0.18,0.12){283}{\line(1,0){0.18}}
\linethickness{0.8mm}
\linethickness{0.3mm}
\put(45,10){\line(0,1){35}}
\linethickness{0.3mm}
\multiput(110,90)(0.18,-0.12){163}{\line(1,0){0.18}}
\linethickness{0.8mm}
\multiput(70,15)(0.12,0.16){125}{\line(0,1){0.16}}
\linethickness{0.8mm}
\put(95,10){\line(0,1){20}}
\linethickness{0.8mm}
\multiput(105,35)(0.12,-0.12){125}{\line(1,0){0.12}}
\put(10,50){\makebox(0,0)[cc]{$p_i$}}

\put(40,15){\makebox(0,0)[cc]{$k_1$}}

\put(65,50){\makebox(0,0)[cc]{$p_i+k_1$}}

\put(150,85){\makebox(0,0)[cc]{$p_j$}}

\put(105,85){\makebox(0,0)[cc]{$k_2$}}

\put(130,52){\makebox(0,0)[cc]{$p_j+k_2$}}

\put(95,45){\makebox(0,0)[cc]{$\Gamma$}}

\put(105,15){\makebox(0,0)[cc]{$\cdot$}}

\put(108,18){\makebox(0,0)[cc]{$\cdot$}}

\end{picture}

}

\def\figsoftthreetree{

\def\JPicScale{0.6}
\ifx\JPicScale\undefined\def\JPicScale{1}\fi
\unitlength \JPicScale mm
\begin{picture}(205,80)(0,0)

\linethickness{0.3mm}
\put(180.32,44.75){\line(0,1){0.5}}
\multiput(180.31,45.75)(0.02,-0.5){1}{\line(0,-1){0.5}}
\multiput(180.27,46.25)(0.03,-0.5){1}{\line(0,-1){0.5}}
\multiput(180.22,46.75)(0.05,-0.5){1}{\line(0,-1){0.5}}
\multiput(180.16,47.25)(0.07,-0.5){1}{\line(0,-1){0.5}}
\multiput(180.08,47.74)(0.08,-0.49){1}{\line(0,-1){0.49}}
\multiput(179.98,48.24)(0.1,-0.49){1}{\line(0,-1){0.49}}
\multiput(179.87,48.72)(0.11,-0.49){1}{\line(0,-1){0.49}}
\multiput(179.74,49.21)(0.13,-0.48){1}{\line(0,-1){0.48}}
\multiput(179.59,49.69)(0.15,-0.48){1}{\line(0,-1){0.48}}
\multiput(179.43,50.16)(0.16,-0.47){1}{\line(0,-1){0.47}}
\multiput(179.25,50.63)(0.18,-0.47){1}{\line(0,-1){0.47}}
\multiput(179.06,51.1)(0.1,-0.23){2}{\line(0,-1){0.23}}
\multiput(178.85,51.55)(0.1,-0.23){2}{\line(0,-1){0.23}}
\multiput(178.63,52)(0.11,-0.22){2}{\line(0,-1){0.22}}
\multiput(178.4,52.44)(0.12,-0.22){2}{\line(0,-1){0.22}}
\multiput(178.14,52.88)(0.13,-0.22){2}{\line(0,-1){0.22}}
\multiput(177.88,53.3)(0.13,-0.21){2}{\line(0,-1){0.21}}
\multiput(177.6,53.72)(0.14,-0.21){2}{\line(0,-1){0.21}}
\multiput(177.31,54.13)(0.15,-0.2){2}{\line(0,-1){0.2}}
\multiput(177,54.53)(0.1,-0.13){3}{\line(0,-1){0.13}}
\multiput(176.69,54.91)(0.11,-0.13){3}{\line(0,-1){0.13}}
\multiput(176.36,55.29)(0.11,-0.13){3}{\line(0,-1){0.13}}
\multiput(176.01,55.66)(0.11,-0.12){3}{\line(0,-1){0.12}}
\multiput(175.66,56.01)(0.12,-0.12){3}{\line(0,-1){0.12}}
\multiput(175.29,56.36)(0.12,-0.11){3}{\line(1,0){0.12}}
\multiput(174.91,56.69)(0.13,-0.11){3}{\line(1,0){0.13}}
\multiput(174.53,57)(0.13,-0.11){3}{\line(1,0){0.13}}
\multiput(174.13,57.31)(0.13,-0.1){3}{\line(1,0){0.13}}
\multiput(173.72,57.6)(0.2,-0.15){2}{\line(1,0){0.2}}
\multiput(173.3,57.88)(0.21,-0.14){2}{\line(1,0){0.21}}
\multiput(172.88,58.14)(0.21,-0.13){2}{\line(1,0){0.21}}
\multiput(172.44,58.4)(0.22,-0.13){2}{\line(1,0){0.22}}
\multiput(172,58.63)(0.22,-0.12){2}{\line(1,0){0.22}}
\multiput(171.55,58.85)(0.22,-0.11){2}{\line(1,0){0.22}}
\multiput(171.1,59.06)(0.23,-0.1){2}{\line(1,0){0.23}}
\multiput(170.63,59.25)(0.23,-0.1){2}{\line(1,0){0.23}}
\multiput(170.16,59.43)(0.47,-0.18){1}{\line(1,0){0.47}}
\multiput(169.69,59.59)(0.47,-0.16){1}{\line(1,0){0.47}}
\multiput(169.21,59.74)(0.48,-0.15){1}{\line(1,0){0.48}}
\multiput(168.72,59.87)(0.48,-0.13){1}{\line(1,0){0.48}}
\multiput(168.24,59.98)(0.49,-0.11){1}{\line(1,0){0.49}}
\multiput(167.74,60.08)(0.49,-0.1){1}{\line(1,0){0.49}}
\multiput(167.25,60.16)(0.49,-0.08){1}{\line(1,0){0.49}}
\multiput(166.75,60.22)(0.5,-0.07){1}{\line(1,0){0.5}}
\multiput(166.25,60.27)(0.5,-0.05){1}{\line(1,0){0.5}}
\multiput(165.75,60.31)(0.5,-0.03){1}{\line(1,0){0.5}}
\multiput(165.25,60.32)(0.5,-0.02){1}{\line(1,0){0.5}}
\put(164.75,60.32){\line(1,0){0.5}}
\multiput(164.25,60.31)(0.5,0.02){1}{\line(1,0){0.5}}
\multiput(163.75,60.27)(0.5,0.03){1}{\line(1,0){0.5}}
\multiput(163.25,60.22)(0.5,0.05){1}{\line(1,0){0.5}}
\multiput(162.75,60.16)(0.5,0.07){1}{\line(1,0){0.5}}
\multiput(162.26,60.08)(0.49,0.08){1}{\line(1,0){0.49}}
\multiput(161.76,59.98)(0.49,0.1){1}{\line(1,0){0.49}}
\multiput(161.28,59.87)(0.49,0.11){1}{\line(1,0){0.49}}
\multiput(160.79,59.74)(0.48,0.13){1}{\line(1,0){0.48}}
\multiput(160.31,59.59)(0.48,0.15){1}{\line(1,0){0.48}}
\multiput(159.84,59.43)(0.47,0.16){1}{\line(1,0){0.47}}
\multiput(159.37,59.25)(0.47,0.18){1}{\line(1,0){0.47}}
\multiput(158.9,59.06)(0.23,0.1){2}{\line(1,0){0.23}}
\multiput(158.45,58.85)(0.23,0.1){2}{\line(1,0){0.23}}
\multiput(158,58.63)(0.22,0.11){2}{\line(1,0){0.22}}
\multiput(157.56,58.4)(0.22,0.12){2}{\line(1,0){0.22}}
\multiput(157.12,58.14)(0.22,0.13){2}{\line(1,0){0.22}}
\multiput(156.7,57.88)(0.21,0.13){2}{\line(1,0){0.21}}
\multiput(156.28,57.6)(0.21,0.14){2}{\line(1,0){0.21}}
\multiput(155.87,57.31)(0.2,0.15){2}{\line(1,0){0.2}}
\multiput(155.47,57)(0.13,0.1){3}{\line(1,0){0.13}}
\multiput(155.09,56.69)(0.13,0.11){3}{\line(1,0){0.13}}
\multiput(154.71,56.36)(0.13,0.11){3}{\line(1,0){0.13}}
\multiput(154.34,56.01)(0.12,0.11){3}{\line(1,0){0.12}}
\multiput(153.99,55.66)(0.12,0.12){3}{\line(0,1){0.12}}
\multiput(153.64,55.29)(0.11,0.12){3}{\line(0,1){0.12}}
\multiput(153.31,54.91)(0.11,0.13){3}{\line(0,1){0.13}}
\multiput(153,54.53)(0.11,0.13){3}{\line(0,1){0.13}}
\multiput(152.69,54.13)(0.1,0.13){3}{\line(0,1){0.13}}
\multiput(152.4,53.72)(0.15,0.2){2}{\line(0,1){0.2}}
\multiput(152.12,53.3)(0.14,0.21){2}{\line(0,1){0.21}}
\multiput(151.86,52.88)(0.13,0.21){2}{\line(0,1){0.21}}
\multiput(151.6,52.44)(0.13,0.22){2}{\line(0,1){0.22}}
\multiput(151.37,52)(0.12,0.22){2}{\line(0,1){0.22}}
\multiput(151.15,51.55)(0.11,0.22){2}{\line(0,1){0.22}}
\multiput(150.94,51.1)(0.1,0.23){2}{\line(0,1){0.23}}
\multiput(150.75,50.63)(0.1,0.23){2}{\line(0,1){0.23}}
\multiput(150.57,50.16)(0.18,0.47){1}{\line(0,1){0.47}}
\multiput(150.41,49.69)(0.16,0.47){1}{\line(0,1){0.47}}
\multiput(150.26,49.21)(0.15,0.48){1}{\line(0,1){0.48}}
\multiput(150.13,48.72)(0.13,0.48){1}{\line(0,1){0.48}}
\multiput(150.02,48.24)(0.11,0.49){1}{\line(0,1){0.49}}
\multiput(149.92,47.74)(0.1,0.49){1}{\line(0,1){0.49}}
\multiput(149.84,47.25)(0.08,0.49){1}{\line(0,1){0.49}}
\multiput(149.78,46.75)(0.07,0.5){1}{\line(0,1){0.5}}
\multiput(149.73,46.25)(0.05,0.5){1}{\line(0,1){0.5}}
\multiput(149.69,45.75)(0.03,0.5){1}{\line(0,1){0.5}}
\multiput(149.68,45.25)(0.02,0.5){1}{\line(0,1){0.5}}
\put(149.68,44.75){\line(0,1){0.5}}
\multiput(149.68,44.75)(0.02,-0.5){1}{\line(0,-1){0.5}}
\multiput(149.69,44.25)(0.03,-0.5){1}{\line(0,-1){0.5}}
\multiput(149.73,43.75)(0.05,-0.5){1}{\line(0,-1){0.5}}
\multiput(149.78,43.25)(0.07,-0.5){1}{\line(0,-1){0.5}}
\multiput(149.84,42.75)(0.08,-0.49){1}{\line(0,-1){0.49}}
\multiput(149.92,42.26)(0.1,-0.49){1}{\line(0,-1){0.49}}
\multiput(150.02,41.76)(0.11,-0.49){1}{\line(0,-1){0.49}}
\multiput(150.13,41.28)(0.13,-0.48){1}{\line(0,-1){0.48}}
\multiput(150.26,40.79)(0.15,-0.48){1}{\line(0,-1){0.48}}
\multiput(150.41,40.31)(0.16,-0.47){1}{\line(0,-1){0.47}}
\multiput(150.57,39.84)(0.18,-0.47){1}{\line(0,-1){0.47}}
\multiput(150.75,39.37)(0.1,-0.23){2}{\line(0,-1){0.23}}
\multiput(150.94,38.9)(0.1,-0.23){2}{\line(0,-1){0.23}}
\multiput(151.15,38.45)(0.11,-0.22){2}{\line(0,-1){0.22}}
\multiput(151.37,38)(0.12,-0.22){2}{\line(0,-1){0.22}}
\multiput(151.6,37.56)(0.13,-0.22){2}{\line(0,-1){0.22}}
\multiput(151.86,37.12)(0.13,-0.21){2}{\line(0,-1){0.21}}
\multiput(152.12,36.7)(0.14,-0.21){2}{\line(0,-1){0.21}}
\multiput(152.4,36.28)(0.15,-0.2){2}{\line(0,-1){0.2}}
\multiput(152.69,35.87)(0.1,-0.13){3}{\line(0,-1){0.13}}
\multiput(153,35.47)(0.11,-0.13){3}{\line(0,-1){0.13}}
\multiput(153.31,35.09)(0.11,-0.13){3}{\line(0,-1){0.13}}
\multiput(153.64,34.71)(0.11,-0.12){3}{\line(0,-1){0.12}}
\multiput(153.99,34.34)(0.12,-0.12){3}{\line(0,-1){0.12}}
\multiput(154.34,33.99)(0.12,-0.11){3}{\line(1,0){0.12}}
\multiput(154.71,33.64)(0.13,-0.11){3}{\line(1,0){0.13}}
\multiput(155.09,33.31)(0.13,-0.11){3}{\line(1,0){0.13}}
\multiput(155.47,33)(0.13,-0.1){3}{\line(1,0){0.13}}
\multiput(155.87,32.69)(0.2,-0.15){2}{\line(1,0){0.2}}
\multiput(156.28,32.4)(0.21,-0.14){2}{\line(1,0){0.21}}
\multiput(156.7,32.12)(0.21,-0.13){2}{\line(1,0){0.21}}
\multiput(157.12,31.86)(0.22,-0.13){2}{\line(1,0){0.22}}
\multiput(157.56,31.6)(0.22,-0.12){2}{\line(1,0){0.22}}
\multiput(158,31.37)(0.22,-0.11){2}{\line(1,0){0.22}}
\multiput(158.45,31.15)(0.23,-0.1){2}{\line(1,0){0.23}}
\multiput(158.9,30.94)(0.23,-0.1){2}{\line(1,0){0.23}}
\multiput(159.37,30.75)(0.47,-0.18){1}{\line(1,0){0.47}}
\multiput(159.84,30.57)(0.47,-0.16){1}{\line(1,0){0.47}}
\multiput(160.31,30.41)(0.48,-0.15){1}{\line(1,0){0.48}}
\multiput(160.79,30.26)(0.48,-0.13){1}{\line(1,0){0.48}}
\multiput(161.28,30.13)(0.49,-0.11){1}{\line(1,0){0.49}}
\multiput(161.76,30.02)(0.49,-0.1){1}{\line(1,0){0.49}}
\multiput(162.26,29.92)(0.49,-0.08){1}{\line(1,0){0.49}}
\multiput(162.75,29.84)(0.5,-0.07){1}{\line(1,0){0.5}}
\multiput(163.25,29.78)(0.5,-0.05){1}{\line(1,0){0.5}}
\multiput(163.75,29.73)(0.5,-0.03){1}{\line(1,0){0.5}}
\multiput(164.25,29.69)(0.5,-0.02){1}{\line(1,0){0.5}}
\put(164.75,29.68){\line(1,0){0.5}}
\multiput(165.25,29.68)(0.5,0.02){1}{\line(1,0){0.5}}
\multiput(165.75,29.69)(0.5,0.03){1}{\line(1,0){0.5}}
\multiput(166.25,29.73)(0.5,0.05){1}{\line(1,0){0.5}}
\multiput(166.75,29.78)(0.5,0.07){1}{\line(1,0){0.5}}
\multiput(167.25,29.84)(0.49,0.08){1}{\line(1,0){0.49}}
\multiput(167.74,29.92)(0.49,0.1){1}{\line(1,0){0.49}}
\multiput(168.24,30.02)(0.49,0.11){1}{\line(1,0){0.49}}
\multiput(168.72,30.13)(0.48,0.13){1}{\line(1,0){0.48}}
\multiput(169.21,30.26)(0.48,0.15){1}{\line(1,0){0.48}}
\multiput(169.69,30.41)(0.47,0.16){1}{\line(1,0){0.47}}
\multiput(170.16,30.57)(0.47,0.18){1}{\line(1,0){0.47}}
\multiput(170.63,30.75)(0.23,0.1){2}{\line(1,0){0.23}}
\multiput(171.1,30.94)(0.23,0.1){2}{\line(1,0){0.23}}
\multiput(171.55,31.15)(0.22,0.11){2}{\line(1,0){0.22}}
\multiput(172,31.37)(0.22,0.12){2}{\line(1,0){0.22}}
\multiput(172.44,31.6)(0.22,0.13){2}{\line(1,0){0.22}}
\multiput(172.88,31.86)(0.21,0.13){2}{\line(1,0){0.21}}
\multiput(173.3,32.12)(0.21,0.14){2}{\line(1,0){0.21}}
\multiput(173.72,32.4)(0.2,0.15){2}{\line(1,0){0.2}}
\multiput(174.13,32.69)(0.13,0.1){3}{\line(1,0){0.13}}
\multiput(174.53,33)(0.13,0.11){3}{\line(1,0){0.13}}
\multiput(174.91,33.31)(0.13,0.11){3}{\line(1,0){0.13}}
\multiput(175.29,33.64)(0.12,0.11){3}{\line(1,0){0.12}}
\multiput(175.66,33.99)(0.12,0.12){3}{\line(0,1){0.12}}
\multiput(176.01,34.34)(0.11,0.12){3}{\line(0,1){0.12}}
\multiput(176.36,34.71)(0.11,0.13){3}{\line(0,1){0.13}}
\multiput(176.69,35.09)(0.11,0.13){3}{\line(0,1){0.13}}
\multiput(177,35.47)(0.1,0.13){3}{\line(0,1){0.13}}
\multiput(177.31,35.87)(0.15,0.2){2}{\line(0,1){0.2}}
\multiput(177.6,36.28)(0.14,0.21){2}{\line(0,1){0.21}}
\multiput(177.88,36.7)(0.13,0.21){2}{\line(0,1){0.21}}
\multiput(178.14,37.12)(0.13,0.22){2}{\line(0,1){0.22}}
\multiput(178.4,37.56)(0.12,0.22){2}{\line(0,1){0.22}}
\multiput(178.63,38)(0.11,0.22){2}{\line(0,1){0.22}}
\multiput(178.85,38.45)(0.1,0.23){2}{\line(0,1){0.23}}
\multiput(179.06,38.9)(0.1,0.23){2}{\line(0,1){0.23}}
\multiput(179.25,39.37)(0.18,0.47){1}{\line(0,1){0.47}}
\multiput(179.43,39.84)(0.16,0.47){1}{\line(0,1){0.47}}
\multiput(179.59,40.31)(0.15,0.48){1}{\line(0,1){0.48}}
\multiput(179.74,40.79)(0.13,0.48){1}{\line(0,1){0.48}}
\multiput(179.87,41.28)(0.11,0.49){1}{\line(0,1){0.49}}
\multiput(179.98,41.76)(0.1,0.49){1}{\line(0,1){0.49}}
\multiput(180.08,42.26)(0.08,0.49){1}{\line(0,1){0.49}}
\multiput(180.16,42.75)(0.07,0.5){1}{\line(0,1){0.5}}
\multiput(180.22,43.25)(0.05,0.5){1}{\line(0,1){0.5}}
\multiput(180.27,43.75)(0.03,0.5){1}{\line(0,1){0.5}}
\multiput(180.31,44.25)(0.02,0.5){1}{\line(0,1){0.5}}

\linethickness{0.8mm}
\put(0,45){\line(1,0){40}}
\linethickness{0.8mm}
\put(40,45){\line(1,0){50}}
\linethickness{0.8mm}
\put(90,45){\line(1,0){60}}
\linethickness{0.3mm}
\put(25,10){\line(0,1){35}}
\linethickness{0.3mm}
\put(75,10){\line(0,1){35}}
\linethickness{0.8mm}
\multiput(170,60)(0.15,0.12){167}{\line(1,0){0.15}}
\linethickness{0.8mm}
\multiput(175,35)(0.18,-0.12){167}{\line(1,0){0.18}}
\linethickness{0.8mm}
\multiput(175,55)(0.3,0.12){83}{\line(1,0){0.3}}
\put(5,50){\makebox(0,0)[cc]{$p_i$}}

\put(30,15){\makebox(0,0)[cc]{$k_1$}}

\put(80,15){\makebox(0,0)[cc]{$k_2$}}

\put(50,50){\makebox(0,0)[cc]{$p_i+k_1$}}

\put(120,50){\makebox(0,0)[cc]{$p_i+k_1+k_2$}}

\put(190,45){\makebox(0,0)[cc]{$\cdot$}}

\put(185,35){\makebox(0,0)[cc]{$\cdot$}}

\put(165,45){\makebox(0,0)[cc]{$\Gamma$}}

\end{picture}

}

\def\figsoftonefieldtree{

\def\JPicScale{0.6}
\ifx\JPicScale\undefined\def\JPicScale{1}\fi
\unitlength \JPicScale mm
\begin{picture}(135,85)(0,0)

\linethickness{0.3mm}
\put(110.36,49.75){\line(0,1){0.5}}
\multiput(110.34,50.75)(0.02,-0.5){1}{\line(0,-1){0.5}}
\multiput(110.31,51.26)(0.03,-0.5){1}{\line(0,-1){0.5}}
\multiput(110.26,51.76)(0.05,-0.5){1}{\line(0,-1){0.5}}
\multiput(110.2,52.25)(0.07,-0.5){1}{\line(0,-1){0.5}}
\multiput(110.11,52.75)(0.08,-0.5){1}{\line(0,-1){0.5}}
\multiput(110.02,53.24)(0.1,-0.49){1}{\line(0,-1){0.49}}
\multiput(109.9,53.73)(0.11,-0.49){1}{\line(0,-1){0.49}}
\multiput(109.77,54.22)(0.13,-0.49){1}{\line(0,-1){0.49}}
\multiput(109.62,54.7)(0.15,-0.48){1}{\line(0,-1){0.48}}
\multiput(109.46,55.18)(0.16,-0.48){1}{\line(0,-1){0.48}}
\multiput(109.29,55.65)(0.18,-0.47){1}{\line(0,-1){0.47}}
\multiput(109.09,56.11)(0.1,-0.23){2}{\line(0,-1){0.23}}
\multiput(108.89,56.57)(0.1,-0.23){2}{\line(0,-1){0.23}}
\multiput(108.66,57.02)(0.11,-0.23){2}{\line(0,-1){0.23}}
\multiput(108.43,57.46)(0.12,-0.22){2}{\line(0,-1){0.22}}
\multiput(108.18,57.9)(0.13,-0.22){2}{\line(0,-1){0.22}}
\multiput(107.91,58.32)(0.13,-0.21){2}{\line(0,-1){0.21}}
\multiput(107.63,58.74)(0.14,-0.21){2}{\line(0,-1){0.21}}
\multiput(107.34,59.15)(0.15,-0.2){2}{\line(0,-1){0.2}}
\multiput(107.03,59.55)(0.1,-0.13){3}{\line(0,-1){0.13}}
\multiput(106.71,59.94)(0.11,-0.13){3}{\line(0,-1){0.13}}
\multiput(106.38,60.32)(0.11,-0.13){3}{\line(0,-1){0.13}}
\multiput(106.04,60.68)(0.11,-0.12){3}{\line(0,-1){0.12}}
\multiput(105.68,61.04)(0.12,-0.12){3}{\line(1,0){0.12}}
\multiput(105.32,61.38)(0.12,-0.11){3}{\line(1,0){0.12}}
\multiput(104.94,61.71)(0.13,-0.11){3}{\line(1,0){0.13}}
\multiput(104.55,62.03)(0.13,-0.11){3}{\line(1,0){0.13}}
\multiput(104.15,62.34)(0.13,-0.1){3}{\line(1,0){0.13}}
\multiput(103.74,62.63)(0.2,-0.15){2}{\line(1,0){0.2}}
\multiput(103.32,62.91)(0.21,-0.14){2}{\line(1,0){0.21}}
\multiput(102.9,63.18)(0.21,-0.13){2}{\line(1,0){0.21}}
\multiput(102.46,63.43)(0.22,-0.13){2}{\line(1,0){0.22}}
\multiput(102.02,63.66)(0.22,-0.12){2}{\line(1,0){0.22}}
\multiput(101.57,63.89)(0.23,-0.11){2}{\line(1,0){0.23}}
\multiput(101.11,64.09)(0.23,-0.1){2}{\line(1,0){0.23}}
\multiput(100.65,64.29)(0.23,-0.1){2}{\line(1,0){0.23}}
\multiput(100.18,64.46)(0.47,-0.18){1}{\line(1,0){0.47}}
\multiput(99.7,64.62)(0.48,-0.16){1}{\line(1,0){0.48}}
\multiput(99.22,64.77)(0.48,-0.15){1}{\line(1,0){0.48}}
\multiput(98.73,64.9)(0.49,-0.13){1}{\line(1,0){0.49}}
\multiput(98.24,65.02)(0.49,-0.11){1}{\line(1,0){0.49}}
\multiput(97.75,65.11)(0.49,-0.1){1}{\line(1,0){0.49}}
\multiput(97.25,65.2)(0.5,-0.08){1}{\line(1,0){0.5}}
\multiput(96.76,65.26)(0.5,-0.07){1}{\line(1,0){0.5}}
\multiput(96.26,65.31)(0.5,-0.05){1}{\line(1,0){0.5}}
\multiput(95.75,65.34)(0.5,-0.03){1}{\line(1,0){0.5}}
\multiput(95.25,65.36)(0.5,-0.02){1}{\line(1,0){0.5}}
\put(94.75,65.36){\line(1,0){0.5}}
\multiput(94.25,65.34)(0.5,0.02){1}{\line(1,0){0.5}}
\multiput(93.74,65.31)(0.5,0.03){1}{\line(1,0){0.5}}
\multiput(93.24,65.26)(0.5,0.05){1}{\line(1,0){0.5}}
\multiput(92.75,65.2)(0.5,0.07){1}{\line(1,0){0.5}}
\multiput(92.25,65.11)(0.5,0.08){1}{\line(1,0){0.5}}
\multiput(91.76,65.02)(0.49,0.1){1}{\line(1,0){0.49}}
\multiput(91.27,64.9)(0.49,0.11){1}{\line(1,0){0.49}}
\multiput(90.78,64.77)(0.49,0.13){1}{\line(1,0){0.49}}
\multiput(90.3,64.62)(0.48,0.15){1}{\line(1,0){0.48}}
\multiput(89.82,64.46)(0.48,0.16){1}{\line(1,0){0.48}}
\multiput(89.35,64.29)(0.47,0.18){1}{\line(1,0){0.47}}
\multiput(88.89,64.09)(0.23,0.1){2}{\line(1,0){0.23}}
\multiput(88.43,63.89)(0.23,0.1){2}{\line(1,0){0.23}}
\multiput(87.98,63.66)(0.23,0.11){2}{\line(1,0){0.23}}
\multiput(87.54,63.43)(0.22,0.12){2}{\line(1,0){0.22}}
\multiput(87.1,63.18)(0.22,0.13){2}{\line(1,0){0.22}}
\multiput(86.68,62.91)(0.21,0.13){2}{\line(1,0){0.21}}
\multiput(86.26,62.63)(0.21,0.14){2}{\line(1,0){0.21}}
\multiput(85.85,62.34)(0.2,0.15){2}{\line(1,0){0.2}}
\multiput(85.45,62.03)(0.13,0.1){3}{\line(1,0){0.13}}
\multiput(85.06,61.71)(0.13,0.11){3}{\line(1,0){0.13}}
\multiput(84.68,61.38)(0.13,0.11){3}{\line(1,0){0.13}}
\multiput(84.32,61.04)(0.12,0.11){3}{\line(1,0){0.12}}
\multiput(83.96,60.68)(0.12,0.12){3}{\line(0,1){0.12}}
\multiput(83.62,60.32)(0.11,0.12){3}{\line(0,1){0.12}}
\multiput(83.29,59.94)(0.11,0.13){3}{\line(0,1){0.13}}
\multiput(82.97,59.55)(0.11,0.13){3}{\line(0,1){0.13}}
\multiput(82.66,59.15)(0.1,0.13){3}{\line(0,1){0.13}}
\multiput(82.37,58.74)(0.15,0.2){2}{\line(0,1){0.2}}
\multiput(82.09,58.32)(0.14,0.21){2}{\line(0,1){0.21}}
\multiput(81.82,57.9)(0.13,0.21){2}{\line(0,1){0.21}}
\multiput(81.57,57.46)(0.13,0.22){2}{\line(0,1){0.22}}
\multiput(81.34,57.02)(0.12,0.22){2}{\line(0,1){0.22}}
\multiput(81.11,56.57)(0.11,0.23){2}{\line(0,1){0.23}}
\multiput(80.91,56.11)(0.1,0.23){2}{\line(0,1){0.23}}
\multiput(80.71,55.65)(0.1,0.23){2}{\line(0,1){0.23}}
\multiput(80.54,55.18)(0.18,0.47){1}{\line(0,1){0.47}}
\multiput(80.38,54.7)(0.16,0.48){1}{\line(0,1){0.48}}
\multiput(80.23,54.22)(0.15,0.48){1}{\line(0,1){0.48}}
\multiput(80.1,53.73)(0.13,0.49){1}{\line(0,1){0.49}}
\multiput(79.98,53.24)(0.11,0.49){1}{\line(0,1){0.49}}
\multiput(79.89,52.75)(0.1,0.49){1}{\line(0,1){0.49}}
\multiput(79.8,52.25)(0.08,0.5){1}{\line(0,1){0.5}}
\multiput(79.74,51.76)(0.07,0.5){1}{\line(0,1){0.5}}
\multiput(79.69,51.26)(0.05,0.5){1}{\line(0,1){0.5}}
\multiput(79.66,50.75)(0.03,0.5){1}{\line(0,1){0.5}}
\multiput(79.64,50.25)(0.02,0.5){1}{\line(0,1){0.5}}
\put(79.64,49.75){\line(0,1){0.5}}
\multiput(79.64,49.75)(0.02,-0.5){1}{\line(0,-1){0.5}}
\multiput(79.66,49.25)(0.03,-0.5){1}{\line(0,-1){0.5}}
\multiput(79.69,48.74)(0.05,-0.5){1}{\line(0,-1){0.5}}
\multiput(79.74,48.24)(0.07,-0.5){1}{\line(0,-1){0.5}}
\multiput(79.8,47.75)(0.08,-0.5){1}{\line(0,-1){0.5}}
\multiput(79.89,47.25)(0.1,-0.49){1}{\line(0,-1){0.49}}
\multiput(79.98,46.76)(0.11,-0.49){1}{\line(0,-1){0.49}}
\multiput(80.1,46.27)(0.13,-0.49){1}{\line(0,-1){0.49}}
\multiput(80.23,45.78)(0.15,-0.48){1}{\line(0,-1){0.48}}
\multiput(80.38,45.3)(0.16,-0.48){1}{\line(0,-1){0.48}}
\multiput(80.54,44.82)(0.18,-0.47){1}{\line(0,-1){0.47}}
\multiput(80.71,44.35)(0.1,-0.23){2}{\line(0,-1){0.23}}
\multiput(80.91,43.89)(0.1,-0.23){2}{\line(0,-1){0.23}}
\multiput(81.11,43.43)(0.11,-0.23){2}{\line(0,-1){0.23}}
\multiput(81.34,42.98)(0.12,-0.22){2}{\line(0,-1){0.22}}
\multiput(81.57,42.54)(0.13,-0.22){2}{\line(0,-1){0.22}}
\multiput(81.82,42.1)(0.13,-0.21){2}{\line(0,-1){0.21}}
\multiput(82.09,41.68)(0.14,-0.21){2}{\line(0,-1){0.21}}
\multiput(82.37,41.26)(0.15,-0.2){2}{\line(0,-1){0.2}}
\multiput(82.66,40.85)(0.1,-0.13){3}{\line(0,-1){0.13}}
\multiput(82.97,40.45)(0.11,-0.13){3}{\line(0,-1){0.13}}
\multiput(83.29,40.06)(0.11,-0.13){3}{\line(0,-1){0.13}}
\multiput(83.62,39.68)(0.11,-0.12){3}{\line(0,-1){0.12}}
\multiput(83.96,39.32)(0.12,-0.12){3}{\line(0,-1){0.12}}
\multiput(84.32,38.96)(0.12,-0.11){3}{\line(1,0){0.12}}
\multiput(84.68,38.62)(0.13,-0.11){3}{\line(1,0){0.13}}
\multiput(85.06,38.29)(0.13,-0.11){3}{\line(1,0){0.13}}
\multiput(85.45,37.97)(0.13,-0.1){3}{\line(1,0){0.13}}
\multiput(85.85,37.66)(0.2,-0.15){2}{\line(1,0){0.2}}
\multiput(86.26,37.37)(0.21,-0.14){2}{\line(1,0){0.21}}
\multiput(86.68,37.09)(0.21,-0.13){2}{\line(1,0){0.21}}
\multiput(87.1,36.82)(0.22,-0.13){2}{\line(1,0){0.22}}
\multiput(87.54,36.57)(0.22,-0.12){2}{\line(1,0){0.22}}
\multiput(87.98,36.34)(0.23,-0.11){2}{\line(1,0){0.23}}
\multiput(88.43,36.11)(0.23,-0.1){2}{\line(1,0){0.23}}
\multiput(88.89,35.91)(0.23,-0.1){2}{\line(1,0){0.23}}
\multiput(89.35,35.71)(0.47,-0.18){1}{\line(1,0){0.47}}
\multiput(89.82,35.54)(0.48,-0.16){1}{\line(1,0){0.48}}
\multiput(90.3,35.38)(0.48,-0.15){1}{\line(1,0){0.48}}
\multiput(90.78,35.23)(0.49,-0.13){1}{\line(1,0){0.49}}
\multiput(91.27,35.1)(0.49,-0.11){1}{\line(1,0){0.49}}
\multiput(91.76,34.98)(0.49,-0.1){1}{\line(1,0){0.49}}
\multiput(92.25,34.89)(0.5,-0.08){1}{\line(1,0){0.5}}
\multiput(92.75,34.8)(0.5,-0.07){1}{\line(1,0){0.5}}
\multiput(93.24,34.74)(0.5,-0.05){1}{\line(1,0){0.5}}
\multiput(93.74,34.69)(0.5,-0.03){1}{\line(1,0){0.5}}
\multiput(94.25,34.66)(0.5,-0.02){1}{\line(1,0){0.5}}
\put(94.75,34.64){\line(1,0){0.5}}
\multiput(95.25,34.64)(0.5,0.02){1}{\line(1,0){0.5}}
\multiput(95.75,34.66)(0.5,0.03){1}{\line(1,0){0.5}}
\multiput(96.26,34.69)(0.5,0.05){1}{\line(1,0){0.5}}
\multiput(96.76,34.74)(0.5,0.07){1}{\line(1,0){0.5}}
\multiput(97.25,34.8)(0.5,0.08){1}{\line(1,0){0.5}}
\multiput(97.75,34.89)(0.49,0.1){1}{\line(1,0){0.49}}
\multiput(98.24,34.98)(0.49,0.11){1}{\line(1,0){0.49}}
\multiput(98.73,35.1)(0.49,0.13){1}{\line(1,0){0.49}}
\multiput(99.22,35.23)(0.48,0.15){1}{\line(1,0){0.48}}
\multiput(99.7,35.38)(0.48,0.16){1}{\line(1,0){0.48}}
\multiput(100.18,35.54)(0.47,0.18){1}{\line(1,0){0.47}}
\multiput(100.65,35.71)(0.23,0.1){2}{\line(1,0){0.23}}
\multiput(101.11,35.91)(0.23,0.1){2}{\line(1,0){0.23}}
\multiput(101.57,36.11)(0.23,0.11){2}{\line(1,0){0.23}}
\multiput(102.02,36.34)(0.22,0.12){2}{\line(1,0){0.22}}
\multiput(102.46,36.57)(0.22,0.13){2}{\line(1,0){0.22}}
\multiput(102.9,36.82)(0.21,0.13){2}{\line(1,0){0.21}}
\multiput(103.32,37.09)(0.21,0.14){2}{\line(1,0){0.21}}
\multiput(103.74,37.37)(0.2,0.15){2}{\line(1,0){0.2}}
\multiput(104.15,37.66)(0.13,0.1){3}{\line(1,0){0.13}}
\multiput(104.55,37.97)(0.13,0.11){3}{\line(1,0){0.13}}
\multiput(104.94,38.29)(0.13,0.11){3}{\line(1,0){0.13}}
\multiput(105.32,38.62)(0.12,0.11){3}{\line(1,0){0.12}}
\multiput(105.68,38.96)(0.12,0.12){3}{\line(1,0){0.12}}
\multiput(106.04,39.32)(0.11,0.12){3}{\line(0,1){0.12}}
\multiput(106.38,39.68)(0.11,0.13){3}{\line(0,1){0.13}}
\multiput(106.71,40.06)(0.11,0.13){3}{\line(0,1){0.13}}
\multiput(107.03,40.45)(0.1,0.13){3}{\line(0,1){0.13}}
\multiput(107.34,40.85)(0.15,0.2){2}{\line(0,1){0.2}}
\multiput(107.63,41.26)(0.14,0.21){2}{\line(0,1){0.21}}
\multiput(107.91,41.68)(0.13,0.21){2}{\line(0,1){0.21}}
\multiput(108.18,42.1)(0.13,0.22){2}{\line(0,1){0.22}}
\multiput(108.43,42.54)(0.12,0.22){2}{\line(0,1){0.22}}
\multiput(108.66,42.98)(0.11,0.23){2}{\line(0,1){0.23}}
\multiput(108.89,43.43)(0.1,0.23){2}{\line(0,1){0.23}}
\multiput(109.09,43.89)(0.1,0.23){2}{\line(0,1){0.23}}
\multiput(109.29,44.35)(0.18,0.47){1}{\line(0,1){0.47}}
\multiput(109.46,44.82)(0.16,0.48){1}{\line(0,1){0.48}}
\multiput(109.62,45.3)(0.15,0.48){1}{\line(0,1){0.48}}
\multiput(109.77,45.78)(0.13,0.49){1}{\line(0,1){0.49}}
\multiput(109.9,46.27)(0.11,0.49){1}{\line(0,1){0.49}}
\multiput(110.02,46.76)(0.1,0.49){1}{\line(0,1){0.49}}
\multiput(110.11,47.25)(0.08,0.5){1}{\line(0,1){0.5}}
\multiput(110.2,47.75)(0.07,0.5){1}{\line(0,1){0.5}}
\multiput(110.26,48.24)(0.05,0.5){1}{\line(0,1){0.5}}
\multiput(110.31,48.74)(0.03,0.5){1}{\line(0,1){0.5}}
\multiput(110.34,49.25)(0.02,0.5){1}{\line(0,1){0.5}}

\linethickness{0.8mm}
\put(5,50){\line(1,0){45}}
\linethickness{0.8mm}
\put(50,50){\line(1,0){30}}
\linethickness{0.3mm}
\put(35,10){\line(0,1){40}}
\linethickness{0.8mm}
\multiput(100,65)(0.16,0.12){125}{\line(1,0){0.16}}
\linethickness{0.8mm}
\put(110,50){\line(1,0){25}}
\linethickness{0.8mm}
\multiput(100,35)(0.12,-0.16){125}{\line(0,-1){0.16}}

\put(95,50){\makebox(0,0)[cc]{$\Gamma$}}

\put(0,55){\makebox(0,0)[cc]{$p_i$}}

\put(65,55){\makebox(0,0)[cc]{$p_i+k$}}

\put(40,25){\makebox(0,0)[cc]{$k$}}

\put(110,82){\makebox(0,0)[cc]{$p_1$}}

\put(130,55){\makebox(0,0)[cc]{$p_{i-1}$}}

\put(108,10){\makebox(0,0)[cc]{$p_N$}}

\put(115,25){\makebox(0,0)[cc]{$\cdot$}}

\put(118,28){\makebox(0,0)[cc]{$\cdot$}}

\put(115,65){\makebox(0,0)[cc]{$\cdot$}}

\put(118,62){\makebox(0,0)[cc]{$\cdot$}}

\linethickness{0.8mm}
\multiput(108,42)(0.2,-0.12){125}{\line(1,0){0.2}}

\put(132,35){\makebox(0,0)[cc]{$p_{i+1}$}}

\end{picture}

}

\def\figsoftonefield{

\def\JPicScale{0.6}
\ifx\JPicScale\undefined\def\JPicScale{1}\fi
\unitlength \JPicScale mm

}

\def\figsoftone{

\def\JPicScale{0.6}
\ifx\JPicScale\undefined\def\JPicScale{1}\fi
\unitlength \JPicScale mm

}

\def\figprmod{

\def\JPicScale{0.6}
\ifx\JPicScale\undefined\def\JPicScale{1}\fi
\unitlength \JPicScale mm
\begin{picture}(200,70.51)(0,0)
\linethickness{0.3mm}
\put(160.5,49.75){\line(0,1){0.5}}
\multiput(160.49,50.75)(0.01,-0.5){1}{\line(0,-1){0.5}}
\multiput(160.47,51.26)(0.02,-0.5){1}{\line(0,-1){0.5}}
\multiput(160.43,51.76)(0.04,-0.5){1}{\line(0,-1){0.5}}
\multiput(160.38,52.26)(0.05,-0.5){1}{\line(0,-1){0.5}}
\multiput(160.32,52.76)(0.06,-0.5){1}{\line(0,-1){0.5}}
\multiput(160.25,53.26)(0.07,-0.5){1}{\line(0,-1){0.5}}
\multiput(160.16,53.75)(0.09,-0.5){1}{\line(0,-1){0.5}}
\multiput(160.06,54.25)(0.1,-0.49){1}{\line(0,-1){0.49}}
\multiput(159.95,54.74)(0.11,-0.49){1}{\line(0,-1){0.49}}
\multiput(159.83,55.23)(0.12,-0.49){1}{\line(0,-1){0.49}}
\multiput(159.69,55.71)(0.13,-0.49){1}{\line(0,-1){0.49}}
\multiput(159.55,56.19)(0.15,-0.48){1}{\line(0,-1){0.48}}
\multiput(159.39,56.67)(0.16,-0.48){1}{\line(0,-1){0.48}}
\multiput(159.22,57.14)(0.17,-0.47){1}{\line(0,-1){0.47}}
\multiput(159.04,57.61)(0.09,-0.23){2}{\line(0,-1){0.23}}
\multiput(158.85,58.08)(0.1,-0.23){2}{\line(0,-1){0.23}}
\multiput(158.64,58.54)(0.1,-0.23){2}{\line(0,-1){0.23}}
\multiput(158.43,58.99)(0.11,-0.23){2}{\line(0,-1){0.23}}
\multiput(158.2,59.44)(0.11,-0.22){2}{\line(0,-1){0.22}}
\multiput(157.96,59.89)(0.12,-0.22){2}{\line(0,-1){0.22}}
\multiput(157.72,60.33)(0.12,-0.22){2}{\line(0,-1){0.22}}
\multiput(157.46,60.76)(0.13,-0.22){2}{\line(0,-1){0.22}}
\multiput(157.19,61.18)(0.13,-0.21){2}{\line(0,-1){0.21}}
\multiput(156.91,61.6)(0.14,-0.21){2}{\line(0,-1){0.21}}
\multiput(156.62,62.01)(0.14,-0.21){2}{\line(0,-1){0.21}}
\multiput(156.32,62.42)(0.15,-0.2){2}{\line(0,-1){0.2}}
\multiput(156.01,62.81)(0.1,-0.13){3}{\line(0,-1){0.13}}
\multiput(155.69,63.2)(0.11,-0.13){3}{\line(0,-1){0.13}}
\multiput(155.36,63.58)(0.11,-0.13){3}{\line(0,-1){0.13}}
\multiput(155.02,63.96)(0.11,-0.12){3}{\line(0,-1){0.12}}
\multiput(154.68,64.32)(0.12,-0.12){3}{\line(0,-1){0.12}}
\multiput(154.32,64.68)(0.12,-0.12){3}{\line(0,-1){0.12}}
\multiput(153.96,65.02)(0.12,-0.12){3}{\line(1,0){0.12}}
\multiput(153.58,65.36)(0.12,-0.11){3}{\line(1,0){0.12}}
\multiput(153.2,65.69)(0.13,-0.11){3}{\line(1,0){0.13}}
\multiput(152.81,66.01)(0.13,-0.11){3}{\line(1,0){0.13}}
\multiput(152.42,66.32)(0.13,-0.1){3}{\line(1,0){0.13}}
\multiput(152.01,66.62)(0.2,-0.15){2}{\line(1,0){0.2}}
\multiput(151.6,66.91)(0.21,-0.14){2}{\line(1,0){0.21}}
\multiput(151.18,67.19)(0.21,-0.14){2}{\line(1,0){0.21}}
\multiput(150.76,67.46)(0.21,-0.13){2}{\line(1,0){0.21}}
\multiput(150.33,67.72)(0.22,-0.13){2}{\line(1,0){0.22}}
\multiput(149.89,67.96)(0.22,-0.12){2}{\line(1,0){0.22}}
\multiput(149.44,68.2)(0.22,-0.12){2}{\line(1,0){0.22}}
\multiput(148.99,68.43)(0.22,-0.11){2}{\line(1,0){0.22}}
\multiput(148.54,68.64)(0.23,-0.11){2}{\line(1,0){0.23}}
\multiput(148.08,68.85)(0.23,-0.1){2}{\line(1,0){0.23}}
\multiput(147.61,69.04)(0.23,-0.1){2}{\line(1,0){0.23}}
\multiput(147.14,69.22)(0.23,-0.09){2}{\line(1,0){0.23}}
\multiput(146.67,69.39)(0.47,-0.17){1}{\line(1,0){0.47}}
\multiput(146.19,69.55)(0.48,-0.16){1}{\line(1,0){0.48}}
\multiput(145.71,69.69)(0.48,-0.15){1}{\line(1,0){0.48}}
\multiput(145.23,69.83)(0.49,-0.13){1}{\line(1,0){0.49}}
\multiput(144.74,69.95)(0.49,-0.12){1}{\line(1,0){0.49}}
\multiput(144.25,70.06)(0.49,-0.11){1}{\line(1,0){0.49}}
\multiput(143.75,70.16)(0.49,-0.1){1}{\line(1,0){0.49}}
\multiput(143.26,70.25)(0.5,-0.09){1}{\line(1,0){0.5}}
\multiput(142.76,70.32)(0.5,-0.07){1}{\line(1,0){0.5}}
\multiput(142.26,70.38)(0.5,-0.06){1}{\line(1,0){0.5}}
\multiput(141.76,70.43)(0.5,-0.05){1}{\line(1,0){0.5}}
\multiput(141.26,70.47)(0.5,-0.04){1}{\line(1,0){0.5}}
\multiput(140.75,70.49)(0.5,-0.02){1}{\line(1,0){0.5}}
\multiput(140.25,70.5)(0.5,-0.01){1}{\line(1,0){0.5}}
\put(139.75,70.5){\line(1,0){0.5}}
\multiput(139.25,70.49)(0.5,0.01){1}{\line(1,0){0.5}}
\multiput(138.74,70.47)(0.5,0.02){1}{\line(1,0){0.5}}
\multiput(138.24,70.43)(0.5,0.04){1}{\line(1,0){0.5}}
\multiput(137.74,70.38)(0.5,0.05){1}{\line(1,0){0.5}}
\multiput(137.24,70.32)(0.5,0.06){1}{\line(1,0){0.5}}
\multiput(136.74,70.25)(0.5,0.07){1}{\line(1,0){0.5}}
\multiput(136.25,70.16)(0.5,0.09){1}{\line(1,0){0.5}}
\multiput(135.75,70.06)(0.49,0.1){1}{\line(1,0){0.49}}
\multiput(135.26,69.95)(0.49,0.11){1}{\line(1,0){0.49}}
\multiput(134.77,69.83)(0.49,0.12){1}{\line(1,0){0.49}}
\multiput(134.29,69.69)(0.49,0.13){1}{\line(1,0){0.49}}
\multiput(133.81,69.55)(0.48,0.15){1}{\line(1,0){0.48}}
\multiput(133.33,69.39)(0.48,0.16){1}{\line(1,0){0.48}}
\multiput(132.86,69.22)(0.47,0.17){1}{\line(1,0){0.47}}
\multiput(132.39,69.04)(0.23,0.09){2}{\line(1,0){0.23}}
\multiput(131.92,68.85)(0.23,0.1){2}{\line(1,0){0.23}}
\multiput(131.46,68.64)(0.23,0.1){2}{\line(1,0){0.23}}
\multiput(131.01,68.43)(0.23,0.11){2}{\line(1,0){0.23}}
\multiput(130.56,68.2)(0.22,0.11){2}{\line(1,0){0.22}}
\multiput(130.11,67.96)(0.22,0.12){2}{\line(1,0){0.22}}
\multiput(129.67,67.72)(0.22,0.12){2}{\line(1,0){0.22}}
\multiput(129.24,67.46)(0.22,0.13){2}{\line(1,0){0.22}}
\multiput(128.82,67.19)(0.21,0.13){2}{\line(1,0){0.21}}
\multiput(128.4,66.91)(0.21,0.14){2}{\line(1,0){0.21}}
\multiput(127.99,66.62)(0.21,0.14){2}{\line(1,0){0.21}}
\multiput(127.58,66.32)(0.2,0.15){2}{\line(1,0){0.2}}
\multiput(127.19,66.01)(0.13,0.1){3}{\line(1,0){0.13}}
\multiput(126.8,65.69)(0.13,0.11){3}{\line(1,0){0.13}}
\multiput(126.42,65.36)(0.13,0.11){3}{\line(1,0){0.13}}
\multiput(126.04,65.02)(0.12,0.11){3}{\line(1,0){0.12}}
\multiput(125.68,64.68)(0.12,0.12){3}{\line(1,0){0.12}}
\multiput(125.32,64.32)(0.12,0.12){3}{\line(1,0){0.12}}
\multiput(124.98,63.96)(0.12,0.12){3}{\line(0,1){0.12}}
\multiput(124.64,63.58)(0.11,0.12){3}{\line(0,1){0.12}}
\multiput(124.31,63.2)(0.11,0.13){3}{\line(0,1){0.13}}
\multiput(123.99,62.81)(0.11,0.13){3}{\line(0,1){0.13}}
\multiput(123.68,62.42)(0.1,0.13){3}{\line(0,1){0.13}}
\multiput(123.38,62.01)(0.15,0.2){2}{\line(0,1){0.2}}
\multiput(123.09,61.6)(0.14,0.21){2}{\line(0,1){0.21}}
\multiput(122.81,61.18)(0.14,0.21){2}{\line(0,1){0.21}}
\multiput(122.54,60.76)(0.13,0.21){2}{\line(0,1){0.21}}
\multiput(122.28,60.33)(0.13,0.22){2}{\line(0,1){0.22}}
\multiput(122.04,59.89)(0.12,0.22){2}{\line(0,1){0.22}}
\multiput(121.8,59.44)(0.12,0.22){2}{\line(0,1){0.22}}
\multiput(121.57,58.99)(0.11,0.22){2}{\line(0,1){0.22}}
\multiput(121.36,58.54)(0.11,0.23){2}{\line(0,1){0.23}}
\multiput(121.15,58.08)(0.1,0.23){2}{\line(0,1){0.23}}
\multiput(120.96,57.61)(0.1,0.23){2}{\line(0,1){0.23}}
\multiput(120.78,57.14)(0.09,0.23){2}{\line(0,1){0.23}}
\multiput(120.61,56.67)(0.17,0.47){1}{\line(0,1){0.47}}
\multiput(120.45,56.19)(0.16,0.48){1}{\line(0,1){0.48}}
\multiput(120.31,55.71)(0.15,0.48){1}{\line(0,1){0.48}}
\multiput(120.17,55.23)(0.13,0.49){1}{\line(0,1){0.49}}
\multiput(120.05,54.74)(0.12,0.49){1}{\line(0,1){0.49}}
\multiput(119.94,54.25)(0.11,0.49){1}{\line(0,1){0.49}}
\multiput(119.84,53.75)(0.1,0.49){1}{\line(0,1){0.49}}
\multiput(119.75,53.26)(0.09,0.5){1}{\line(0,1){0.5}}
\multiput(119.68,52.76)(0.07,0.5){1}{\line(0,1){0.5}}
\multiput(119.62,52.26)(0.06,0.5){1}{\line(0,1){0.5}}
\multiput(119.57,51.76)(0.05,0.5){1}{\line(0,1){0.5}}
\multiput(119.53,51.26)(0.04,0.5){1}{\line(0,1){0.5}}
\multiput(119.51,50.75)(0.02,0.5){1}{\line(0,1){0.5}}
\multiput(119.5,50.25)(0.01,0.5){1}{\line(0,1){0.5}}
\put(119.5,49.75){\line(0,1){0.5}}
\multiput(119.5,49.75)(0.01,-0.5){1}{\line(0,-1){0.5}}
\multiput(119.51,49.25)(0.02,-0.5){1}{\line(0,-1){0.5}}
\multiput(119.53,48.74)(0.04,-0.5){1}{\line(0,-1){0.5}}
\multiput(119.57,48.24)(0.05,-0.5){1}{\line(0,-1){0.5}}
\multiput(119.62,47.74)(0.06,-0.5){1}{\line(0,-1){0.5}}
\multiput(119.68,47.24)(0.07,-0.5){1}{\line(0,-1){0.5}}
\multiput(119.75,46.74)(0.09,-0.5){1}{\line(0,-1){0.5}}
\multiput(119.84,46.25)(0.1,-0.49){1}{\line(0,-1){0.49}}
\multiput(119.94,45.75)(0.11,-0.49){1}{\line(0,-1){0.49}}
\multiput(120.05,45.26)(0.12,-0.49){1}{\line(0,-1){0.49}}
\multiput(120.17,44.77)(0.13,-0.49){1}{\line(0,-1){0.49}}
\multiput(120.31,44.29)(0.15,-0.48){1}{\line(0,-1){0.48}}
\multiput(120.45,43.81)(0.16,-0.48){1}{\line(0,-1){0.48}}
\multiput(120.61,43.33)(0.17,-0.47){1}{\line(0,-1){0.47}}
\multiput(120.78,42.86)(0.09,-0.23){2}{\line(0,-1){0.23}}
\multiput(120.96,42.39)(0.1,-0.23){2}{\line(0,-1){0.23}}
\multiput(121.15,41.92)(0.1,-0.23){2}{\line(0,-1){0.23}}
\multiput(121.36,41.46)(0.11,-0.23){2}{\line(0,-1){0.23}}
\multiput(121.57,41.01)(0.11,-0.22){2}{\line(0,-1){0.22}}
\multiput(121.8,40.56)(0.12,-0.22){2}{\line(0,-1){0.22}}
\multiput(122.04,40.11)(0.12,-0.22){2}{\line(0,-1){0.22}}
\multiput(122.28,39.67)(0.13,-0.22){2}{\line(0,-1){0.22}}
\multiput(122.54,39.24)(0.13,-0.21){2}{\line(0,-1){0.21}}
\multiput(122.81,38.82)(0.14,-0.21){2}{\line(0,-1){0.21}}
\multiput(123.09,38.4)(0.14,-0.21){2}{\line(0,-1){0.21}}
\multiput(123.38,37.99)(0.15,-0.2){2}{\line(0,-1){0.2}}
\multiput(123.68,37.58)(0.1,-0.13){3}{\line(0,-1){0.13}}
\multiput(123.99,37.19)(0.11,-0.13){3}{\line(0,-1){0.13}}
\multiput(124.31,36.8)(0.11,-0.13){3}{\line(0,-1){0.13}}
\multiput(124.64,36.42)(0.11,-0.12){3}{\line(0,-1){0.12}}
\multiput(124.98,36.04)(0.12,-0.12){3}{\line(0,-1){0.12}}
\multiput(125.32,35.68)(0.12,-0.12){3}{\line(1,0){0.12}}
\multiput(125.68,35.32)(0.12,-0.12){3}{\line(1,0){0.12}}
\multiput(126.04,34.98)(0.12,-0.11){3}{\line(1,0){0.12}}
\multiput(126.42,34.64)(0.13,-0.11){3}{\line(1,0){0.13}}
\multiput(126.8,34.31)(0.13,-0.11){3}{\line(1,0){0.13}}
\multiput(127.19,33.99)(0.13,-0.1){3}{\line(1,0){0.13}}
\multiput(127.58,33.68)(0.2,-0.15){2}{\line(1,0){0.2}}
\multiput(127.99,33.38)(0.21,-0.14){2}{\line(1,0){0.21}}
\multiput(128.4,33.09)(0.21,-0.14){2}{\line(1,0){0.21}}
\multiput(128.82,32.81)(0.21,-0.13){2}{\line(1,0){0.21}}
\multiput(129.24,32.54)(0.22,-0.13){2}{\line(1,0){0.22}}
\multiput(129.67,32.28)(0.22,-0.12){2}{\line(1,0){0.22}}
\multiput(130.11,32.04)(0.22,-0.12){2}{\line(1,0){0.22}}
\multiput(130.56,31.8)(0.22,-0.11){2}{\line(1,0){0.22}}
\multiput(131.01,31.57)(0.23,-0.11){2}{\line(1,0){0.23}}
\multiput(131.46,31.36)(0.23,-0.1){2}{\line(1,0){0.23}}
\multiput(131.92,31.15)(0.23,-0.1){2}{\line(1,0){0.23}}
\multiput(132.39,30.96)(0.23,-0.09){2}{\line(1,0){0.23}}
\multiput(132.86,30.78)(0.47,-0.17){1}{\line(1,0){0.47}}
\multiput(133.33,30.61)(0.48,-0.16){1}{\line(1,0){0.48}}
\multiput(133.81,30.45)(0.48,-0.15){1}{\line(1,0){0.48}}
\multiput(134.29,30.31)(0.49,-0.13){1}{\line(1,0){0.49}}
\multiput(134.77,30.17)(0.49,-0.12){1}{\line(1,0){0.49}}
\multiput(135.26,30.05)(0.49,-0.11){1}{\line(1,0){0.49}}
\multiput(135.75,29.94)(0.49,-0.1){1}{\line(1,0){0.49}}
\multiput(136.25,29.84)(0.5,-0.09){1}{\line(1,0){0.5}}
\multiput(136.74,29.75)(0.5,-0.07){1}{\line(1,0){0.5}}
\multiput(137.24,29.68)(0.5,-0.06){1}{\line(1,0){0.5}}
\multiput(137.74,29.62)(0.5,-0.05){1}{\line(1,0){0.5}}
\multiput(138.24,29.57)(0.5,-0.04){1}{\line(1,0){0.5}}
\multiput(138.74,29.53)(0.5,-0.02){1}{\line(1,0){0.5}}
\multiput(139.25,29.51)(0.5,-0.01){1}{\line(1,0){0.5}}
\put(139.75,29.5){\line(1,0){0.5}}
\multiput(140.25,29.5)(0.5,0.01){1}{\line(1,0){0.5}}
\multiput(140.75,29.51)(0.5,0.02){1}{\line(1,0){0.5}}
\multiput(141.26,29.53)(0.5,0.04){1}{\line(1,0){0.5}}
\multiput(141.76,29.57)(0.5,0.05){1}{\line(1,0){0.5}}
\multiput(142.26,29.62)(0.5,0.06){1}{\line(1,0){0.5}}
\multiput(142.76,29.68)(0.5,0.07){1}{\line(1,0){0.5}}
\multiput(143.26,29.75)(0.5,0.09){1}{\line(1,0){0.5}}
\multiput(143.75,29.84)(0.49,0.1){1}{\line(1,0){0.49}}
\multiput(144.25,29.94)(0.49,0.11){1}{\line(1,0){0.49}}
\multiput(144.74,30.05)(0.49,0.12){1}{\line(1,0){0.49}}
\multiput(145.23,30.17)(0.49,0.13){1}{\line(1,0){0.49}}
\multiput(145.71,30.31)(0.48,0.15){1}{\line(1,0){0.48}}
\multiput(146.19,30.45)(0.48,0.16){1}{\line(1,0){0.48}}
\multiput(146.67,30.61)(0.47,0.17){1}{\line(1,0){0.47}}
\multiput(147.14,30.78)(0.23,0.09){2}{\line(1,0){0.23}}
\multiput(147.61,30.96)(0.23,0.1){2}{\line(1,0){0.23}}
\multiput(148.08,31.15)(0.23,0.1){2}{\line(1,0){0.23}}
\multiput(148.54,31.36)(0.23,0.11){2}{\line(1,0){0.23}}
\multiput(148.99,31.57)(0.22,0.11){2}{\line(1,0){0.22}}
\multiput(149.44,31.8)(0.22,0.12){2}{\line(1,0){0.22}}
\multiput(149.89,32.04)(0.22,0.12){2}{\line(1,0){0.22}}
\multiput(150.33,32.28)(0.22,0.13){2}{\line(1,0){0.22}}
\multiput(150.76,32.54)(0.21,0.13){2}{\line(1,0){0.21}}
\multiput(151.18,32.81)(0.21,0.14){2}{\line(1,0){0.21}}
\multiput(151.6,33.09)(0.21,0.14){2}{\line(1,0){0.21}}
\multiput(152.01,33.38)(0.2,0.15){2}{\line(1,0){0.2}}
\multiput(152.42,33.68)(0.13,0.1){3}{\line(1,0){0.13}}
\multiput(152.81,33.99)(0.13,0.11){3}{\line(1,0){0.13}}
\multiput(153.2,34.31)(0.13,0.11){3}{\line(1,0){0.13}}
\multiput(153.58,34.64)(0.12,0.11){3}{\line(1,0){0.12}}
\multiput(153.96,34.98)(0.12,0.12){3}{\line(1,0){0.12}}
\multiput(154.32,35.32)(0.12,0.12){3}{\line(0,1){0.12}}
\multiput(154.68,35.68)(0.12,0.12){3}{\line(0,1){0.12}}
\multiput(155.02,36.04)(0.11,0.12){3}{\line(0,1){0.12}}
\multiput(155.36,36.42)(0.11,0.13){3}{\line(0,1){0.13}}
\multiput(155.69,36.8)(0.11,0.13){3}{\line(0,1){0.13}}
\multiput(156.01,37.19)(0.1,0.13){3}{\line(0,1){0.13}}
\multiput(156.32,37.58)(0.15,0.2){2}{\line(0,1){0.2}}
\multiput(156.62,37.99)(0.14,0.21){2}{\line(0,1){0.21}}
\multiput(156.91,38.4)(0.14,0.21){2}{\line(0,1){0.21}}
\multiput(157.19,38.82)(0.13,0.21){2}{\line(0,1){0.21}}
\multiput(157.46,39.24)(0.13,0.22){2}{\line(0,1){0.22}}
\multiput(157.72,39.67)(0.12,0.22){2}{\line(0,1){0.22}}
\multiput(157.96,40.11)(0.12,0.22){2}{\line(0,1){0.22}}
\multiput(158.2,40.56)(0.11,0.22){2}{\line(0,1){0.22}}
\multiput(158.43,41.01)(0.11,0.23){2}{\line(0,1){0.23}}
\multiput(158.64,41.46)(0.1,0.23){2}{\line(0,1){0.23}}
\multiput(158.85,41.92)(0.1,0.23){2}{\line(0,1){0.23}}
\multiput(159.04,42.39)(0.09,0.23){2}{\line(0,1){0.23}}
\multiput(159.22,42.86)(0.17,0.47){1}{\line(0,1){0.47}}
\multiput(159.39,43.33)(0.16,0.48){1}{\line(0,1){0.48}}
\multiput(159.55,43.81)(0.15,0.48){1}{\line(0,1){0.48}}
\multiput(159.69,44.29)(0.13,0.49){1}{\line(0,1){0.49}}
\multiput(159.83,44.77)(0.12,0.49){1}{\line(0,1){0.49}}
\multiput(159.95,45.26)(0.11,0.49){1}{\line(0,1){0.49}}
\multiput(160.06,45.75)(0.1,0.49){1}{\line(0,1){0.49}}
\multiput(160.16,46.25)(0.09,0.5){1}{\line(0,1){0.5}}
\multiput(160.25,46.74)(0.07,0.5){1}{\line(0,1){0.5}}
\multiput(160.32,47.24)(0.06,0.5){1}{\line(0,1){0.5}}
\multiput(160.38,47.74)(0.05,0.5){1}{\line(0,1){0.5}}
\multiput(160.43,48.24)(0.04,0.5){1}{\line(0,1){0.5}}
\multiput(160.47,48.74)(0.02,0.5){1}{\line(0,1){0.5}}
\multiput(160.49,49.25)(0.01,0.5){1}{\line(0,1){0.5}}

\linethickness{1mm}
\put(85,50){\line(1,0){35}}
\linethickness{1mm}
\put(160,50){\line(1,0){40}}
\linethickness{1mm}
\put(10,50){\line(1,0){50}}
\linethickness{0.3mm}
\put(140,20){\line(0,1){10}}
\put(140,15){\makebox(0,0)[cc]{$\Phi$}}

\put(65,50){\makebox(0,0)[cc]{$ ~ \qquad  + \quad   \lambda$}}

\put(35,45){\makebox(0,0)[cc]{$i\Delta_F$}}

\put(100,45){\makebox(0,0)[cc]{$i\Delta_F$}}

\put(185,45){\makebox(0,0)[cc]{$i\Delta_F$}}

\put(140,50){\makebox(0,0)[cc]{$\Gamma$}}

\end{picture}

}

\def\figwtgam{
\def\JPicScale{0.6}
\ifx\JPicScale\undefined\def\JPicScale{1}\fi
\unitlength \JPicScale mm

}

\def\figsoftthreefieldtree{

\def\JPicScale{0.5}
\ifx\JPicScale\undefined\def\JPicScale{1}\fi
\unitlength \JPicScale mm
\begin{picture}(135,90)(0,0)
\linethickness{0.3mm}
\put(105.03,48.5){\line(0,1){0.5}}
\multiput(105.02,49.5)(0.01,-0.5){1}{\line(0,-1){0.5}}
\multiput(105,50)(0.02,-0.5){1}{\line(0,-1){0.5}}
\multiput(104.97,50.49)(0.03,-0.5){1}{\line(0,-1){0.5}}
\multiput(104.92,50.99)(0.04,-0.5){1}{\line(0,-1){0.5}}
\multiput(104.87,51.49)(0.06,-0.5){1}{\line(0,-1){0.5}}
\multiput(104.8,51.98)(0.07,-0.49){1}{\line(0,-1){0.49}}
\multiput(104.73,52.47)(0.08,-0.49){1}{\line(0,-1){0.49}}
\multiput(104.64,52.96)(0.09,-0.49){1}{\line(0,-1){0.49}}
\multiput(104.54,53.45)(0.1,-0.49){1}{\line(0,-1){0.49}}
\multiput(104.43,53.94)(0.11,-0.49){1}{\line(0,-1){0.49}}
\multiput(104.31,54.42)(0.12,-0.48){1}{\line(0,-1){0.48}}
\multiput(104.18,54.9)(0.13,-0.48){1}{\line(0,-1){0.48}}
\multiput(104.04,55.38)(0.14,-0.48){1}{\line(0,-1){0.48}}
\multiput(103.89,55.86)(0.15,-0.47){1}{\line(0,-1){0.47}}
\multiput(103.72,56.33)(0.16,-0.47){1}{\line(0,-1){0.47}}
\multiput(103.55,56.79)(0.17,-0.47){1}{\line(0,-1){0.47}}
\multiput(103.37,57.26)(0.09,-0.23){2}{\line(0,-1){0.23}}
\multiput(103.17,57.72)(0.1,-0.23){2}{\line(0,-1){0.23}}
\multiput(102.97,58.17)(0.1,-0.23){2}{\line(0,-1){0.23}}
\multiput(102.76,58.62)(0.11,-0.23){2}{\line(0,-1){0.23}}
\multiput(102.53,59.07)(0.11,-0.22){2}{\line(0,-1){0.22}}
\multiput(102.3,59.51)(0.12,-0.22){2}{\line(0,-1){0.22}}
\multiput(102.06,59.95)(0.12,-0.22){2}{\line(0,-1){0.22}}
\multiput(101.8,60.38)(0.13,-0.21){2}{\line(0,-1){0.21}}
\multiput(101.54,60.8)(0.13,-0.21){2}{\line(0,-1){0.21}}
\multiput(101.27,61.22)(0.14,-0.21){2}{\line(0,-1){0.21}}
\multiput(100.99,61.63)(0.14,-0.21){2}{\line(0,-1){0.21}}
\multiput(100.7,62.04)(0.14,-0.2){2}{\line(0,-1){0.2}}
\multiput(100.4,62.44)(0.15,-0.2){2}{\line(0,-1){0.2}}
\multiput(100.1,62.83)(0.1,-0.13){3}{\line(0,-1){0.13}}
\multiput(99.78,63.21)(0.11,-0.13){3}{\line(0,-1){0.13}}
\multiput(99.46,63.59)(0.11,-0.13){3}{\line(0,-1){0.13}}
\multiput(99.12,63.96)(0.11,-0.12){3}{\line(0,-1){0.12}}
\multiput(98.78,64.33)(0.11,-0.12){3}{\line(0,-1){0.12}}
\multiput(98.43,64.68)(0.12,-0.12){3}{\line(0,-1){0.12}}
\multiput(98.08,65.03)(0.12,-0.12){3}{\line(1,0){0.12}}
\multiput(97.71,65.37)(0.12,-0.11){3}{\line(1,0){0.12}}
\multiput(97.34,65.71)(0.12,-0.11){3}{\line(1,0){0.12}}
\multiput(96.96,66.03)(0.13,-0.11){3}{\line(1,0){0.13}}
\multiput(96.58,66.35)(0.13,-0.11){3}{\line(1,0){0.13}}
\multiput(96.19,66.65)(0.13,-0.1){3}{\line(1,0){0.13}}
\multiput(95.79,66.95)(0.2,-0.15){2}{\line(1,0){0.2}}
\multiput(95.38,67.24)(0.2,-0.14){2}{\line(1,0){0.2}}
\multiput(94.97,67.52)(0.21,-0.14){2}{\line(1,0){0.21}}
\multiput(94.55,67.79)(0.21,-0.14){2}{\line(1,0){0.21}}
\multiput(94.13,68.05)(0.21,-0.13){2}{\line(1,0){0.21}}
\multiput(93.7,68.31)(0.21,-0.13){2}{\line(1,0){0.21}}
\multiput(93.26,68.55)(0.22,-0.12){2}{\line(1,0){0.22}}
\multiput(92.82,68.78)(0.22,-0.12){2}{\line(1,0){0.22}}
\multiput(92.37,69.01)(0.22,-0.11){2}{\line(1,0){0.22}}
\multiput(91.92,69.22)(0.23,-0.11){2}{\line(1,0){0.23}}
\multiput(91.47,69.42)(0.23,-0.1){2}{\line(1,0){0.23}}
\multiput(91.01,69.62)(0.23,-0.1){2}{\line(1,0){0.23}}
\multiput(90.54,69.8)(0.23,-0.09){2}{\line(1,0){0.23}}
\multiput(90.08,69.97)(0.47,-0.17){1}{\line(1,0){0.47}}
\multiput(89.61,70.14)(0.47,-0.16){1}{\line(1,0){0.47}}
\multiput(89.13,70.29)(0.47,-0.15){1}{\line(1,0){0.47}}
\multiput(88.65,70.43)(0.48,-0.14){1}{\line(1,0){0.48}}
\multiput(88.17,70.56)(0.48,-0.13){1}{\line(1,0){0.48}}
\multiput(87.69,70.68)(0.48,-0.12){1}{\line(1,0){0.48}}
\multiput(87.2,70.79)(0.49,-0.11){1}{\line(1,0){0.49}}
\multiput(86.71,70.89)(0.49,-0.1){1}{\line(1,0){0.49}}
\multiput(86.22,70.98)(0.49,-0.09){1}{\line(1,0){0.49}}
\multiput(85.73,71.05)(0.49,-0.08){1}{\line(1,0){0.49}}
\multiput(85.24,71.12)(0.49,-0.07){1}{\line(1,0){0.49}}
\multiput(84.74,71.17)(0.5,-0.06){1}{\line(1,0){0.5}}
\multiput(84.24,71.22)(0.5,-0.04){1}{\line(1,0){0.5}}
\multiput(83.75,71.25)(0.5,-0.03){1}{\line(1,0){0.5}}
\multiput(83.25,71.27)(0.5,-0.02){1}{\line(1,0){0.5}}
\multiput(82.75,71.28)(0.5,-0.01){1}{\line(1,0){0.5}}
\put(82.25,71.28){\line(1,0){0.5}}
\multiput(81.75,71.27)(0.5,0.01){1}{\line(1,0){0.5}}
\multiput(81.25,71.25)(0.5,0.02){1}{\line(1,0){0.5}}
\multiput(80.76,71.22)(0.5,0.03){1}{\line(1,0){0.5}}
\multiput(80.26,71.17)(0.5,0.04){1}{\line(1,0){0.5}}
\multiput(79.76,71.12)(0.5,0.06){1}{\line(1,0){0.5}}
\multiput(79.27,71.05)(0.49,0.07){1}{\line(1,0){0.49}}
\multiput(78.78,70.98)(0.49,0.08){1}{\line(1,0){0.49}}
\multiput(78.29,70.89)(0.49,0.09){1}{\line(1,0){0.49}}
\multiput(77.8,70.79)(0.49,0.1){1}{\line(1,0){0.49}}
\multiput(77.31,70.68)(0.49,0.11){1}{\line(1,0){0.49}}
\multiput(76.83,70.56)(0.48,0.12){1}{\line(1,0){0.48}}
\multiput(76.35,70.43)(0.48,0.13){1}{\line(1,0){0.48}}
\multiput(75.87,70.29)(0.48,0.14){1}{\line(1,0){0.48}}
\multiput(75.39,70.14)(0.47,0.15){1}{\line(1,0){0.47}}
\multiput(74.92,69.97)(0.47,0.16){1}{\line(1,0){0.47}}
\multiput(74.46,69.8)(0.47,0.17){1}{\line(1,0){0.47}}
\multiput(73.99,69.62)(0.23,0.09){2}{\line(1,0){0.23}}
\multiput(73.53,69.42)(0.23,0.1){2}{\line(1,0){0.23}}
\multiput(73.08,69.22)(0.23,0.1){2}{\line(1,0){0.23}}
\multiput(72.63,69.01)(0.23,0.11){2}{\line(1,0){0.23}}
\multiput(72.18,68.78)(0.22,0.11){2}{\line(1,0){0.22}}
\multiput(71.74,68.55)(0.22,0.12){2}{\line(1,0){0.22}}
\multiput(71.3,68.31)(0.22,0.12){2}{\line(1,0){0.22}}
\multiput(70.87,68.05)(0.21,0.13){2}{\line(1,0){0.21}}
\multiput(70.45,67.79)(0.21,0.13){2}{\line(1,0){0.21}}
\multiput(70.03,67.52)(0.21,0.14){2}{\line(1,0){0.21}}
\multiput(69.62,67.24)(0.21,0.14){2}{\line(1,0){0.21}}
\multiput(69.21,66.95)(0.2,0.14){2}{\line(1,0){0.2}}
\multiput(68.81,66.65)(0.2,0.15){2}{\line(1,0){0.2}}
\multiput(68.42,66.35)(0.13,0.1){3}{\line(1,0){0.13}}
\multiput(68.04,66.03)(0.13,0.11){3}{\line(1,0){0.13}}
\multiput(67.66,65.71)(0.13,0.11){3}{\line(1,0){0.13}}
\multiput(67.29,65.37)(0.12,0.11){3}{\line(1,0){0.12}}
\multiput(66.92,65.03)(0.12,0.11){3}{\line(1,0){0.12}}
\multiput(66.57,64.68)(0.12,0.12){3}{\line(1,0){0.12}}
\multiput(66.22,64.33)(0.12,0.12){3}{\line(0,1){0.12}}
\multiput(65.88,63.96)(0.11,0.12){3}{\line(0,1){0.12}}
\multiput(65.54,63.59)(0.11,0.12){3}{\line(0,1){0.12}}
\multiput(65.22,63.21)(0.11,0.13){3}{\line(0,1){0.13}}
\multiput(64.9,62.83)(0.11,0.13){3}{\line(0,1){0.13}}
\multiput(64.6,62.44)(0.1,0.13){3}{\line(0,1){0.13}}
\multiput(64.3,62.04)(0.15,0.2){2}{\line(0,1){0.2}}
\multiput(64.01,61.63)(0.14,0.2){2}{\line(0,1){0.2}}
\multiput(63.73,61.22)(0.14,0.21){2}{\line(0,1){0.21}}
\multiput(63.46,60.8)(0.14,0.21){2}{\line(0,1){0.21}}
\multiput(63.2,60.38)(0.13,0.21){2}{\line(0,1){0.21}}
\multiput(62.94,59.95)(0.13,0.21){2}{\line(0,1){0.21}}
\multiput(62.7,59.51)(0.12,0.22){2}{\line(0,1){0.22}}
\multiput(62.47,59.07)(0.12,0.22){2}{\line(0,1){0.22}}
\multiput(62.24,58.62)(0.11,0.22){2}{\line(0,1){0.22}}
\multiput(62.03,58.17)(0.11,0.23){2}{\line(0,1){0.23}}
\multiput(61.83,57.72)(0.1,0.23){2}{\line(0,1){0.23}}
\multiput(61.63,57.26)(0.1,0.23){2}{\line(0,1){0.23}}
\multiput(61.45,56.79)(0.09,0.23){2}{\line(0,1){0.23}}
\multiput(61.28,56.33)(0.17,0.47){1}{\line(0,1){0.47}}
\multiput(61.11,55.86)(0.16,0.47){1}{\line(0,1){0.47}}
\multiput(60.96,55.38)(0.15,0.47){1}{\line(0,1){0.47}}
\multiput(60.82,54.9)(0.14,0.48){1}{\line(0,1){0.48}}
\multiput(60.69,54.42)(0.13,0.48){1}{\line(0,1){0.48}}
\multiput(60.57,53.94)(0.12,0.48){1}{\line(0,1){0.48}}
\multiput(60.46,53.45)(0.11,0.49){1}{\line(0,1){0.49}}
\multiput(60.36,52.96)(0.1,0.49){1}{\line(0,1){0.49}}
\multiput(60.27,52.47)(0.09,0.49){1}{\line(0,1){0.49}}
\multiput(60.2,51.98)(0.08,0.49){1}{\line(0,1){0.49}}
\multiput(60.13,51.49)(0.07,0.49){1}{\line(0,1){0.49}}
\multiput(60.08,50.99)(0.06,0.5){1}{\line(0,1){0.5}}
\multiput(60.03,50.49)(0.04,0.5){1}{\line(0,1){0.5}}
\multiput(60,50)(0.03,0.5){1}{\line(0,1){0.5}}
\multiput(59.98,49.5)(0.02,0.5){1}{\line(0,1){0.5}}
\multiput(59.97,49)(0.01,0.5){1}{\line(0,1){0.5}}
\put(59.97,48.5){\line(0,1){0.5}}
\multiput(59.97,48.5)(0.01,-0.5){1}{\line(0,-1){0.5}}
\multiput(59.98,48)(0.02,-0.5){1}{\line(0,-1){0.5}}
\multiput(60,47.5)(0.03,-0.5){1}{\line(0,-1){0.5}}
\multiput(60.03,47.01)(0.04,-0.5){1}{\line(0,-1){0.5}}
\multiput(60.08,46.51)(0.06,-0.5){1}{\line(0,-1){0.5}}
\multiput(60.13,46.01)(0.07,-0.49){1}{\line(0,-1){0.49}}
\multiput(60.2,45.52)(0.08,-0.49){1}{\line(0,-1){0.49}}
\multiput(60.27,45.03)(0.09,-0.49){1}{\line(0,-1){0.49}}
\multiput(60.36,44.54)(0.1,-0.49){1}{\line(0,-1){0.49}}
\multiput(60.46,44.05)(0.11,-0.49){1}{\line(0,-1){0.49}}
\multiput(60.57,43.56)(0.12,-0.48){1}{\line(0,-1){0.48}}
\multiput(60.69,43.08)(0.13,-0.48){1}{\line(0,-1){0.48}}
\multiput(60.82,42.6)(0.14,-0.48){1}{\line(0,-1){0.48}}
\multiput(60.96,42.12)(0.15,-0.47){1}{\line(0,-1){0.47}}
\multiput(61.11,41.64)(0.16,-0.47){1}{\line(0,-1){0.47}}
\multiput(61.28,41.17)(0.17,-0.47){1}{\line(0,-1){0.47}}
\multiput(61.45,40.71)(0.09,-0.23){2}{\line(0,-1){0.23}}
\multiput(61.63,40.24)(0.1,-0.23){2}{\line(0,-1){0.23}}
\multiput(61.83,39.78)(0.1,-0.23){2}{\line(0,-1){0.23}}
\multiput(62.03,39.33)(0.11,-0.23){2}{\line(0,-1){0.23}}
\multiput(62.24,38.88)(0.11,-0.22){2}{\line(0,-1){0.22}}
\multiput(62.47,38.43)(0.12,-0.22){2}{\line(0,-1){0.22}}
\multiput(62.7,37.99)(0.12,-0.22){2}{\line(0,-1){0.22}}
\multiput(62.94,37.55)(0.13,-0.21){2}{\line(0,-1){0.21}}
\multiput(63.2,37.12)(0.13,-0.21){2}{\line(0,-1){0.21}}
\multiput(63.46,36.7)(0.14,-0.21){2}{\line(0,-1){0.21}}
\multiput(63.73,36.28)(0.14,-0.21){2}{\line(0,-1){0.21}}
\multiput(64.01,35.87)(0.14,-0.2){2}{\line(0,-1){0.2}}
\multiput(64.3,35.46)(0.15,-0.2){2}{\line(0,-1){0.2}}
\multiput(64.6,35.06)(0.1,-0.13){3}{\line(0,-1){0.13}}
\multiput(64.9,34.67)(0.11,-0.13){3}{\line(0,-1){0.13}}
\multiput(65.22,34.29)(0.11,-0.13){3}{\line(0,-1){0.13}}
\multiput(65.54,33.91)(0.11,-0.12){3}{\line(0,-1){0.12}}
\multiput(65.88,33.54)(0.11,-0.12){3}{\line(0,-1){0.12}}
\multiput(66.22,33.17)(0.12,-0.12){3}{\line(0,-1){0.12}}
\multiput(66.57,32.82)(0.12,-0.12){3}{\line(1,0){0.12}}
\multiput(66.92,32.47)(0.12,-0.11){3}{\line(1,0){0.12}}
\multiput(67.29,32.13)(0.12,-0.11){3}{\line(1,0){0.12}}
\multiput(67.66,31.79)(0.13,-0.11){3}{\line(1,0){0.13}}
\multiput(68.04,31.47)(0.13,-0.11){3}{\line(1,0){0.13}}
\multiput(68.42,31.15)(0.13,-0.1){3}{\line(1,0){0.13}}
\multiput(68.81,30.85)(0.2,-0.15){2}{\line(1,0){0.2}}
\multiput(69.21,30.55)(0.2,-0.14){2}{\line(1,0){0.2}}
\multiput(69.62,30.26)(0.21,-0.14){2}{\line(1,0){0.21}}
\multiput(70.03,29.98)(0.21,-0.14){2}{\line(1,0){0.21}}
\multiput(70.45,29.71)(0.21,-0.13){2}{\line(1,0){0.21}}
\multiput(70.87,29.45)(0.21,-0.13){2}{\line(1,0){0.21}}
\multiput(71.3,29.19)(0.22,-0.12){2}{\line(1,0){0.22}}
\multiput(71.74,28.95)(0.22,-0.12){2}{\line(1,0){0.22}}
\multiput(72.18,28.72)(0.22,-0.11){2}{\line(1,0){0.22}}
\multiput(72.63,28.49)(0.23,-0.11){2}{\line(1,0){0.23}}
\multiput(73.08,28.28)(0.23,-0.1){2}{\line(1,0){0.23}}
\multiput(73.53,28.08)(0.23,-0.1){2}{\line(1,0){0.23}}
\multiput(73.99,27.88)(0.23,-0.09){2}{\line(1,0){0.23}}
\multiput(74.46,27.7)(0.47,-0.17){1}{\line(1,0){0.47}}
\multiput(74.92,27.53)(0.47,-0.16){1}{\line(1,0){0.47}}
\multiput(75.39,27.36)(0.47,-0.15){1}{\line(1,0){0.47}}
\multiput(75.87,27.21)(0.48,-0.14){1}{\line(1,0){0.48}}
\multiput(76.35,27.07)(0.48,-0.13){1}{\line(1,0){0.48}}
\multiput(76.83,26.94)(0.48,-0.12){1}{\line(1,0){0.48}}
\multiput(77.31,26.82)(0.49,-0.11){1}{\line(1,0){0.49}}
\multiput(77.8,26.71)(0.49,-0.1){1}{\line(1,0){0.49}}
\multiput(78.29,26.61)(0.49,-0.09){1}{\line(1,0){0.49}}
\multiput(78.78,26.52)(0.49,-0.08){1}{\line(1,0){0.49}}
\multiput(79.27,26.45)(0.49,-0.07){1}{\line(1,0){0.49}}
\multiput(79.76,26.38)(0.5,-0.06){1}{\line(1,0){0.5}}
\multiput(80.26,26.33)(0.5,-0.04){1}{\line(1,0){0.5}}
\multiput(80.76,26.28)(0.5,-0.03){1}{\line(1,0){0.5}}
\multiput(81.25,26.25)(0.5,-0.02){1}{\line(1,0){0.5}}
\multiput(81.75,26.23)(0.5,-0.01){1}{\line(1,0){0.5}}
\put(82.25,26.22){\line(1,0){0.5}}
\multiput(82.75,26.22)(0.5,0.01){1}{\line(1,0){0.5}}
\multiput(83.25,26.23)(0.5,0.02){1}{\line(1,0){0.5}}
\multiput(83.75,26.25)(0.5,0.03){1}{\line(1,0){0.5}}
\multiput(84.24,26.28)(0.5,0.04){1}{\line(1,0){0.5}}
\multiput(84.74,26.33)(0.5,0.06){1}{\line(1,0){0.5}}
\multiput(85.24,26.38)(0.49,0.07){1}{\line(1,0){0.49}}
\multiput(85.73,26.45)(0.49,0.08){1}{\line(1,0){0.49}}
\multiput(86.22,26.52)(0.49,0.09){1}{\line(1,0){0.49}}
\multiput(86.71,26.61)(0.49,0.1){1}{\line(1,0){0.49}}
\multiput(87.2,26.71)(0.49,0.11){1}{\line(1,0){0.49}}
\multiput(87.69,26.82)(0.48,0.12){1}{\line(1,0){0.48}}
\multiput(88.17,26.94)(0.48,0.13){1}{\line(1,0){0.48}}
\multiput(88.65,27.07)(0.48,0.14){1}{\line(1,0){0.48}}
\multiput(89.13,27.21)(0.47,0.15){1}{\line(1,0){0.47}}
\multiput(89.61,27.36)(0.47,0.16){1}{\line(1,0){0.47}}
\multiput(90.08,27.53)(0.47,0.17){1}{\line(1,0){0.47}}
\multiput(90.54,27.7)(0.23,0.09){2}{\line(1,0){0.23}}
\multiput(91.01,27.88)(0.23,0.1){2}{\line(1,0){0.23}}
\multiput(91.47,28.08)(0.23,0.1){2}{\line(1,0){0.23}}
\multiput(91.92,28.28)(0.23,0.11){2}{\line(1,0){0.23}}
\multiput(92.37,28.49)(0.22,0.11){2}{\line(1,0){0.22}}
\multiput(92.82,28.72)(0.22,0.12){2}{\line(1,0){0.22}}
\multiput(93.26,28.95)(0.22,0.12){2}{\line(1,0){0.22}}
\multiput(93.7,29.19)(0.21,0.13){2}{\line(1,0){0.21}}
\multiput(94.13,29.45)(0.21,0.13){2}{\line(1,0){0.21}}
\multiput(94.55,29.71)(0.21,0.14){2}{\line(1,0){0.21}}
\multiput(94.97,29.98)(0.21,0.14){2}{\line(1,0){0.21}}
\multiput(95.38,30.26)(0.2,0.14){2}{\line(1,0){0.2}}
\multiput(95.79,30.55)(0.2,0.15){2}{\line(1,0){0.2}}
\multiput(96.19,30.85)(0.13,0.1){3}{\line(1,0){0.13}}
\multiput(96.58,31.15)(0.13,0.11){3}{\line(1,0){0.13}}
\multiput(96.96,31.47)(0.13,0.11){3}{\line(1,0){0.13}}
\multiput(97.34,31.79)(0.12,0.11){3}{\line(1,0){0.12}}
\multiput(97.71,32.13)(0.12,0.11){3}{\line(1,0){0.12}}
\multiput(98.08,32.47)(0.12,0.12){3}{\line(1,0){0.12}}
\multiput(98.43,32.82)(0.12,0.12){3}{\line(0,1){0.12}}
\multiput(98.78,33.17)(0.11,0.12){3}{\line(0,1){0.12}}
\multiput(99.12,33.54)(0.11,0.12){3}{\line(0,1){0.12}}
\multiput(99.46,33.91)(0.11,0.13){3}{\line(0,1){0.13}}
\multiput(99.78,34.29)(0.11,0.13){3}{\line(0,1){0.13}}
\multiput(100.1,34.67)(0.1,0.13){3}{\line(0,1){0.13}}
\multiput(100.4,35.06)(0.15,0.2){2}{\line(0,1){0.2}}
\multiput(100.7,35.46)(0.14,0.2){2}{\line(0,1){0.2}}
\multiput(100.99,35.87)(0.14,0.21){2}{\line(0,1){0.21}}
\multiput(101.27,36.28)(0.14,0.21){2}{\line(0,1){0.21}}
\multiput(101.54,36.7)(0.13,0.21){2}{\line(0,1){0.21}}
\multiput(101.8,37.12)(0.13,0.21){2}{\line(0,1){0.21}}
\multiput(102.06,37.55)(0.12,0.22){2}{\line(0,1){0.22}}
\multiput(102.3,37.99)(0.12,0.22){2}{\line(0,1){0.22}}
\multiput(102.53,38.43)(0.11,0.22){2}{\line(0,1){0.22}}
\multiput(102.76,38.88)(0.11,0.23){2}{\line(0,1){0.23}}
\multiput(102.97,39.33)(0.1,0.23){2}{\line(0,1){0.23}}
\multiput(103.17,39.78)(0.1,0.23){2}{\line(0,1){0.23}}
\multiput(103.37,40.24)(0.09,0.23){2}{\line(0,1){0.23}}
\multiput(103.55,40.71)(0.17,0.47){1}{\line(0,1){0.47}}
\multiput(103.72,41.17)(0.16,0.47){1}{\line(0,1){0.47}}
\multiput(103.89,41.64)(0.15,0.47){1}{\line(0,1){0.47}}
\multiput(104.04,42.12)(0.14,0.48){1}{\line(0,1){0.48}}
\multiput(104.18,42.6)(0.13,0.48){1}{\line(0,1){0.48}}
\multiput(104.31,43.08)(0.12,0.48){1}{\line(0,1){0.48}}
\multiput(104.43,43.56)(0.11,0.49){1}{\line(0,1){0.49}}
\multiput(104.54,44.05)(0.1,0.49){1}{\line(0,1){0.49}}
\multiput(104.64,44.54)(0.09,0.49){1}{\line(0,1){0.49}}
\multiput(104.73,45.03)(0.08,0.49){1}{\line(0,1){0.49}}
\multiput(104.8,45.52)(0.07,0.49){1}{\line(0,1){0.49}}
\multiput(104.87,46.01)(0.06,0.5){1}{\line(0,1){0.5}}
\multiput(104.92,46.51)(0.04,0.5){1}{\line(0,1){0.5}}
\multiput(104.97,47.01)(0.03,0.5){1}{\line(0,1){0.5}}
\multiput(105,47.5)(0.02,0.5){1}{\line(0,1){0.5}}
\multiput(105.02,48)(0.01,0.5){1}{\line(0,1){0.5}}

\linethickness{1mm}
\put(30,50){\line(1,0){30}}
\linethickness{1mm}
\multiput(60,90)(0.12,-0.16){125}{\line(0,-1){0.16}}
\linethickness{1mm}
\multiput(104,55)(0.36,0.12){83}{\line(1,0){0.36}}
\linethickness{0.3mm}
\multiput(95,30)(0.12,-0.16){125}{\line(0,-1){0.16}}
\put(35,55){\makebox(0,0)[cc]{$p_1$}}

\put(72,85){\makebox(0,0)[cc]{$p_2$}}

\put(95,80){\makebox(0,0)[cc]{$\cdot$}}

\put(110,70){\makebox(0,0)[cc]{$\cdot$}}

\put(120,55){\makebox(0,0)[cc]{$p_N$}}

\put(110,20){\makebox(0,0)[cc]{$k$}}

\put(80,50){\makebox(0,0)[cc]{$\wt\Gamma$}}

\end{picture}

}

\def\figsoftthreefield{

\def\JPicScale{0.6}
\ifx\JPicScale\undefined\def\JPicScale{1}\fi
\unitlength \JPicScale mm
\begin{picture}(135,90)(0,0)
\linethickness{0.3mm}
\put(105.03,48.5){\line(0,1){0.5}}
\multiput(105.02,49.5)(0.01,-0.5){1}{\line(0,-1){0.5}}
\multiput(105,50)(0.02,-0.5){1}{\line(0,-1){0.5}}
\multiput(104.97,50.49)(0.03,-0.5){1}{\line(0,-1){0.5}}
\multiput(104.92,50.99)(0.04,-0.5){1}{\line(0,-1){0.5}}
\multiput(104.87,51.49)(0.06,-0.5){1}{\line(0,-1){0.5}}
\multiput(104.8,51.98)(0.07,-0.49){1}{\line(0,-1){0.49}}
\multiput(104.73,52.47)(0.08,-0.49){1}{\line(0,-1){0.49}}
\multiput(104.64,52.96)(0.09,-0.49){1}{\line(0,-1){0.49}}
\multiput(104.54,53.45)(0.1,-0.49){1}{\line(0,-1){0.49}}
\multiput(104.43,53.94)(0.11,-0.49){1}{\line(0,-1){0.49}}
\multiput(104.31,54.42)(0.12,-0.48){1}{\line(0,-1){0.48}}
\multiput(104.18,54.9)(0.13,-0.48){1}{\line(0,-1){0.48}}
\multiput(104.04,55.38)(0.14,-0.48){1}{\line(0,-1){0.48}}
\multiput(103.89,55.86)(0.15,-0.47){1}{\line(0,-1){0.47}}
\multiput(103.72,56.33)(0.16,-0.47){1}{\line(0,-1){0.47}}
\multiput(103.55,56.79)(0.17,-0.47){1}{\line(0,-1){0.47}}
\multiput(103.37,57.26)(0.09,-0.23){2}{\line(0,-1){0.23}}
\multiput(103.17,57.72)(0.1,-0.23){2}{\line(0,-1){0.23}}
\multiput(102.97,58.17)(0.1,-0.23){2}{\line(0,-1){0.23}}
\multiput(102.76,58.62)(0.11,-0.23){2}{\line(0,-1){0.23}}
\multiput(102.53,59.07)(0.11,-0.22){2}{\line(0,-1){0.22}}
\multiput(102.3,59.51)(0.12,-0.22){2}{\line(0,-1){0.22}}
\multiput(102.06,59.95)(0.12,-0.22){2}{\line(0,-1){0.22}}
\multiput(101.8,60.38)(0.13,-0.21){2}{\line(0,-1){0.21}}
\multiput(101.54,60.8)(0.13,-0.21){2}{\line(0,-1){0.21}}
\multiput(101.27,61.22)(0.14,-0.21){2}{\line(0,-1){0.21}}
\multiput(100.99,61.63)(0.14,-0.21){2}{\line(0,-1){0.21}}
\multiput(100.7,62.04)(0.14,-0.2){2}{\line(0,-1){0.2}}
\multiput(100.4,62.44)(0.15,-0.2){2}{\line(0,-1){0.2}}
\multiput(100.1,62.83)(0.1,-0.13){3}{\line(0,-1){0.13}}
\multiput(99.78,63.21)(0.11,-0.13){3}{\line(0,-1){0.13}}
\multiput(99.46,63.59)(0.11,-0.13){3}{\line(0,-1){0.13}}
\multiput(99.12,63.96)(0.11,-0.12){3}{\line(0,-1){0.12}}
\multiput(98.78,64.33)(0.11,-0.12){3}{\line(0,-1){0.12}}
\multiput(98.43,64.68)(0.12,-0.12){3}{\line(0,-1){0.12}}
\multiput(98.08,65.03)(0.12,-0.12){3}{\line(1,0){0.12}}
\multiput(97.71,65.37)(0.12,-0.11){3}{\line(1,0){0.12}}
\multiput(97.34,65.71)(0.12,-0.11){3}{\line(1,0){0.12}}
\multiput(96.96,66.03)(0.13,-0.11){3}{\line(1,0){0.13}}
\multiput(96.58,66.35)(0.13,-0.11){3}{\line(1,0){0.13}}
\multiput(96.19,66.65)(0.13,-0.1){3}{\line(1,0){0.13}}
\multiput(95.79,66.95)(0.2,-0.15){2}{\line(1,0){0.2}}
\multiput(95.38,67.24)(0.2,-0.14){2}{\line(1,0){0.2}}
\multiput(94.97,67.52)(0.21,-0.14){2}{\line(1,0){0.21}}
\multiput(94.55,67.79)(0.21,-0.14){2}{\line(1,0){0.21}}
\multiput(94.13,68.05)(0.21,-0.13){2}{\line(1,0){0.21}}
\multiput(93.7,68.31)(0.21,-0.13){2}{\line(1,0){0.21}}
\multiput(93.26,68.55)(0.22,-0.12){2}{\line(1,0){0.22}}
\multiput(92.82,68.78)(0.22,-0.12){2}{\line(1,0){0.22}}
\multiput(92.37,69.01)(0.22,-0.11){2}{\line(1,0){0.22}}
\multiput(91.92,69.22)(0.23,-0.11){2}{\line(1,0){0.23}}
\multiput(91.47,69.42)(0.23,-0.1){2}{\line(1,0){0.23}}
\multiput(91.01,69.62)(0.23,-0.1){2}{\line(1,0){0.23}}
\multiput(90.54,69.8)(0.23,-0.09){2}{\line(1,0){0.23}}
\multiput(90.08,69.97)(0.47,-0.17){1}{\line(1,0){0.47}}
\multiput(89.61,70.14)(0.47,-0.16){1}{\line(1,0){0.47}}
\multiput(89.13,70.29)(0.47,-0.15){1}{\line(1,0){0.47}}
\multiput(88.65,70.43)(0.48,-0.14){1}{\line(1,0){0.48}}
\multiput(88.17,70.56)(0.48,-0.13){1}{\line(1,0){0.48}}
\multiput(87.69,70.68)(0.48,-0.12){1}{\line(1,0){0.48}}
\multiput(87.2,70.79)(0.49,-0.11){1}{\line(1,0){0.49}}
\multiput(86.71,70.89)(0.49,-0.1){1}{\line(1,0){0.49}}
\multiput(86.22,70.98)(0.49,-0.09){1}{\line(1,0){0.49}}
\multiput(85.73,71.05)(0.49,-0.08){1}{\line(1,0){0.49}}
\multiput(85.24,71.12)(0.49,-0.07){1}{\line(1,0){0.49}}
\multiput(84.74,71.17)(0.5,-0.06){1}{\line(1,0){0.5}}
\multiput(84.24,71.22)(0.5,-0.04){1}{\line(1,0){0.5}}
\multiput(83.75,71.25)(0.5,-0.03){1}{\line(1,0){0.5}}
\multiput(83.25,71.27)(0.5,-0.02){1}{\line(1,0){0.5}}
\multiput(82.75,71.28)(0.5,-0.01){1}{\line(1,0){0.5}}
\put(82.25,71.28){\line(1,0){0.5}}
\multiput(81.75,71.27)(0.5,0.01){1}{\line(1,0){0.5}}
\multiput(81.25,71.25)(0.5,0.02){1}{\line(1,0){0.5}}
\multiput(80.76,71.22)(0.5,0.03){1}{\line(1,0){0.5}}
\multiput(80.26,71.17)(0.5,0.04){1}{\line(1,0){0.5}}
\multiput(79.76,71.12)(0.5,0.06){1}{\line(1,0){0.5}}
\multiput(79.27,71.05)(0.49,0.07){1}{\line(1,0){0.49}}
\multiput(78.78,70.98)(0.49,0.08){1}{\line(1,0){0.49}}
\multiput(78.29,70.89)(0.49,0.09){1}{\line(1,0){0.49}}
\multiput(77.8,70.79)(0.49,0.1){1}{\line(1,0){0.49}}
\multiput(77.31,70.68)(0.49,0.11){1}{\line(1,0){0.49}}
\multiput(76.83,70.56)(0.48,0.12){1}{\line(1,0){0.48}}
\multiput(76.35,70.43)(0.48,0.13){1}{\line(1,0){0.48}}
\multiput(75.87,70.29)(0.48,0.14){1}{\line(1,0){0.48}}
\multiput(75.39,70.14)(0.47,0.15){1}{\line(1,0){0.47}}
\multiput(74.92,69.97)(0.47,0.16){1}{\line(1,0){0.47}}
\multiput(74.46,69.8)(0.47,0.17){1}{\line(1,0){0.47}}
\multiput(73.99,69.62)(0.23,0.09){2}{\line(1,0){0.23}}
\multiput(73.53,69.42)(0.23,0.1){2}{\line(1,0){0.23}}
\multiput(73.08,69.22)(0.23,0.1){2}{\line(1,0){0.23}}
\multiput(72.63,69.01)(0.23,0.11){2}{\line(1,0){0.23}}
\multiput(72.18,68.78)(0.22,0.11){2}{\line(1,0){0.22}}
\multiput(71.74,68.55)(0.22,0.12){2}{\line(1,0){0.22}}
\multiput(71.3,68.31)(0.22,0.12){2}{\line(1,0){0.22}}
\multiput(70.87,68.05)(0.21,0.13){2}{\line(1,0){0.21}}
\multiput(70.45,67.79)(0.21,0.13){2}{\line(1,0){0.21}}
\multiput(70.03,67.52)(0.21,0.14){2}{\line(1,0){0.21}}
\multiput(69.62,67.24)(0.21,0.14){2}{\line(1,0){0.21}}
\multiput(69.21,66.95)(0.2,0.14){2}{\line(1,0){0.2}}
\multiput(68.81,66.65)(0.2,0.15){2}{\line(1,0){0.2}}
\multiput(68.42,66.35)(0.13,0.1){3}{\line(1,0){0.13}}
\multiput(68.04,66.03)(0.13,0.11){3}{\line(1,0){0.13}}
\multiput(67.66,65.71)(0.13,0.11){3}{\line(1,0){0.13}}
\multiput(67.29,65.37)(0.12,0.11){3}{\line(1,0){0.12}}
\multiput(66.92,65.03)(0.12,0.11){3}{\line(1,0){0.12}}
\multiput(66.57,64.68)(0.12,0.12){3}{\line(1,0){0.12}}
\multiput(66.22,64.33)(0.12,0.12){3}{\line(0,1){0.12}}
\multiput(65.88,63.96)(0.11,0.12){3}{\line(0,1){0.12}}
\multiput(65.54,63.59)(0.11,0.12){3}{\line(0,1){0.12}}
\multiput(65.22,63.21)(0.11,0.13){3}{\line(0,1){0.13}}
\multiput(64.9,62.83)(0.11,0.13){3}{\line(0,1){0.13}}
\multiput(64.6,62.44)(0.1,0.13){3}{\line(0,1){0.13}}
\multiput(64.3,62.04)(0.15,0.2){2}{\line(0,1){0.2}}
\multiput(64.01,61.63)(0.14,0.2){2}{\line(0,1){0.2}}
\multiput(63.73,61.22)(0.14,0.21){2}{\line(0,1){0.21}}
\multiput(63.46,60.8)(0.14,0.21){2}{\line(0,1){0.21}}
\multiput(63.2,60.38)(0.13,0.21){2}{\line(0,1){0.21}}
\multiput(62.94,59.95)(0.13,0.21){2}{\line(0,1){0.21}}
\multiput(62.7,59.51)(0.12,0.22){2}{\line(0,1){0.22}}
\multiput(62.47,59.07)(0.12,0.22){2}{\line(0,1){0.22}}
\multiput(62.24,58.62)(0.11,0.22){2}{\line(0,1){0.22}}
\multiput(62.03,58.17)(0.11,0.23){2}{\line(0,1){0.23}}
\multiput(61.83,57.72)(0.1,0.23){2}{\line(0,1){0.23}}
\multiput(61.63,57.26)(0.1,0.23){2}{\line(0,1){0.23}}
\multiput(61.45,56.79)(0.09,0.23){2}{\line(0,1){0.23}}
\multiput(61.28,56.33)(0.17,0.47){1}{\line(0,1){0.47}}
\multiput(61.11,55.86)(0.16,0.47){1}{\line(0,1){0.47}}
\multiput(60.96,55.38)(0.15,0.47){1}{\line(0,1){0.47}}
\multiput(60.82,54.9)(0.14,0.48){1}{\line(0,1){0.48}}
\multiput(60.69,54.42)(0.13,0.48){1}{\line(0,1){0.48}}
\multiput(60.57,53.94)(0.12,0.48){1}{\line(0,1){0.48}}
\multiput(60.46,53.45)(0.11,0.49){1}{\line(0,1){0.49}}
\multiput(60.36,52.96)(0.1,0.49){1}{\line(0,1){0.49}}
\multiput(60.27,52.47)(0.09,0.49){1}{\line(0,1){0.49}}
\multiput(60.2,51.98)(0.08,0.49){1}{\line(0,1){0.49}}
\multiput(60.13,51.49)(0.07,0.49){1}{\line(0,1){0.49}}
\multiput(60.08,50.99)(0.06,0.5){1}{\line(0,1){0.5}}
\multiput(60.03,50.49)(0.04,0.5){1}{\line(0,1){0.5}}
\multiput(60,50)(0.03,0.5){1}{\line(0,1){0.5}}
\multiput(59.98,49.5)(0.02,0.5){1}{\line(0,1){0.5}}
\multiput(59.97,49)(0.01,0.5){1}{\line(0,1){0.5}}
\put(59.97,48.5){\line(0,1){0.5}}
\multiput(59.97,48.5)(0.01,-0.5){1}{\line(0,-1){0.5}}
\multiput(59.98,48)(0.02,-0.5){1}{\line(0,-1){0.5}}
\multiput(60,47.5)(0.03,-0.5){1}{\line(0,-1){0.5}}
\multiput(60.03,47.01)(0.04,-0.5){1}{\line(0,-1){0.5}}
\multiput(60.08,46.51)(0.06,-0.5){1}{\line(0,-1){0.5}}
\multiput(60.13,46.01)(0.07,-0.49){1}{\line(0,-1){0.49}}
\multiput(60.2,45.52)(0.08,-0.49){1}{\line(0,-1){0.49}}
\multiput(60.27,45.03)(0.09,-0.49){1}{\line(0,-1){0.49}}
\multiput(60.36,44.54)(0.1,-0.49){1}{\line(0,-1){0.49}}
\multiput(60.46,44.05)(0.11,-0.49){1}{\line(0,-1){0.49}}
\multiput(60.57,43.56)(0.12,-0.48){1}{\line(0,-1){0.48}}
\multiput(60.69,43.08)(0.13,-0.48){1}{\line(0,-1){0.48}}
\multiput(60.82,42.6)(0.14,-0.48){1}{\line(0,-1){0.48}}
\multiput(60.96,42.12)(0.15,-0.47){1}{\line(0,-1){0.47}}
\multiput(61.11,41.64)(0.16,-0.47){1}{\line(0,-1){0.47}}
\multiput(61.28,41.17)(0.17,-0.47){1}{\line(0,-1){0.47}}
\multiput(61.45,40.71)(0.09,-0.23){2}{\line(0,-1){0.23}}
\multiput(61.63,40.24)(0.1,-0.23){2}{\line(0,-1){0.23}}
\multiput(61.83,39.78)(0.1,-0.23){2}{\line(0,-1){0.23}}
\multiput(62.03,39.33)(0.11,-0.23){2}{\line(0,-1){0.23}}
\multiput(62.24,38.88)(0.11,-0.22){2}{\line(0,-1){0.22}}
\multiput(62.47,38.43)(0.12,-0.22){2}{\line(0,-1){0.22}}
\multiput(62.7,37.99)(0.12,-0.22){2}{\line(0,-1){0.22}}
\multiput(62.94,37.55)(0.13,-0.21){2}{\line(0,-1){0.21}}
\multiput(63.2,37.12)(0.13,-0.21){2}{\line(0,-1){0.21}}
\multiput(63.46,36.7)(0.14,-0.21){2}{\line(0,-1){0.21}}
\multiput(63.73,36.28)(0.14,-0.21){2}{\line(0,-1){0.21}}
\multiput(64.01,35.87)(0.14,-0.2){2}{\line(0,-1){0.2}}
\multiput(64.3,35.46)(0.15,-0.2){2}{\line(0,-1){0.2}}
\multiput(64.6,35.06)(0.1,-0.13){3}{\line(0,-1){0.13}}
\multiput(64.9,34.67)(0.11,-0.13){3}{\line(0,-1){0.13}}
\multiput(65.22,34.29)(0.11,-0.13){3}{\line(0,-1){0.13}}
\multiput(65.54,33.91)(0.11,-0.12){3}{\line(0,-1){0.12}}
\multiput(65.88,33.54)(0.11,-0.12){3}{\line(0,-1){0.12}}
\multiput(66.22,33.17)(0.12,-0.12){3}{\line(0,-1){0.12}}
\multiput(66.57,32.82)(0.12,-0.12){3}{\line(1,0){0.12}}
\multiput(66.92,32.47)(0.12,-0.11){3}{\line(1,0){0.12}}
\multiput(67.29,32.13)(0.12,-0.11){3}{\line(1,0){0.12}}
\multiput(67.66,31.79)(0.13,-0.11){3}{\line(1,0){0.13}}
\multiput(68.04,31.47)(0.13,-0.11){3}{\line(1,0){0.13}}
\multiput(68.42,31.15)(0.13,-0.1){3}{\line(1,0){0.13}}
\multiput(68.81,30.85)(0.2,-0.15){2}{\line(1,0){0.2}}
\multiput(69.21,30.55)(0.2,-0.14){2}{\line(1,0){0.2}}
\multiput(69.62,30.26)(0.21,-0.14){2}{\line(1,0){0.21}}
\multiput(70.03,29.98)(0.21,-0.14){2}{\line(1,0){0.21}}
\multiput(70.45,29.71)(0.21,-0.13){2}{\line(1,0){0.21}}
\multiput(70.87,29.45)(0.21,-0.13){2}{\line(1,0){0.21}}
\multiput(71.3,29.19)(0.22,-0.12){2}{\line(1,0){0.22}}
\multiput(71.74,28.95)(0.22,-0.12){2}{\line(1,0){0.22}}
\multiput(72.18,28.72)(0.22,-0.11){2}{\line(1,0){0.22}}
\multiput(72.63,28.49)(0.23,-0.11){2}{\line(1,0){0.23}}
\multiput(73.08,28.28)(0.23,-0.1){2}{\line(1,0){0.23}}
\multiput(73.53,28.08)(0.23,-0.1){2}{\line(1,0){0.23}}
\multiput(73.99,27.88)(0.23,-0.09){2}{\line(1,0){0.23}}
\multiput(74.46,27.7)(0.47,-0.17){1}{\line(1,0){0.47}}
\multiput(74.92,27.53)(0.47,-0.16){1}{\line(1,0){0.47}}
\multiput(75.39,27.36)(0.47,-0.15){1}{\line(1,0){0.47}}
\multiput(75.87,27.21)(0.48,-0.14){1}{\line(1,0){0.48}}
\multiput(76.35,27.07)(0.48,-0.13){1}{\line(1,0){0.48}}
\multiput(76.83,26.94)(0.48,-0.12){1}{\line(1,0){0.48}}
\multiput(77.31,26.82)(0.49,-0.11){1}{\line(1,0){0.49}}
\multiput(77.8,26.71)(0.49,-0.1){1}{\line(1,0){0.49}}
\multiput(78.29,26.61)(0.49,-0.09){1}{\line(1,0){0.49}}
\multiput(78.78,26.52)(0.49,-0.08){1}{\line(1,0){0.49}}
\multiput(79.27,26.45)(0.49,-0.07){1}{\line(1,0){0.49}}
\multiput(79.76,26.38)(0.5,-0.06){1}{\line(1,0){0.5}}
\multiput(80.26,26.33)(0.5,-0.04){1}{\line(1,0){0.5}}
\multiput(80.76,26.28)(0.5,-0.03){1}{\line(1,0){0.5}}
\multiput(81.25,26.25)(0.5,-0.02){1}{\line(1,0){0.5}}
\multiput(81.75,26.23)(0.5,-0.01){1}{\line(1,0){0.5}}
\put(82.25,26.22){\line(1,0){0.5}}
\multiput(82.75,26.22)(0.5,0.01){1}{\line(1,0){0.5}}
\multiput(83.25,26.23)(0.5,0.02){1}{\line(1,0){0.5}}
\multiput(83.75,26.25)(0.5,0.03){1}{\line(1,0){0.5}}
\multiput(84.24,26.28)(0.5,0.04){1}{\line(1,0){0.5}}
\multiput(84.74,26.33)(0.5,0.06){1}{\line(1,0){0.5}}
\multiput(85.24,26.38)(0.49,0.07){1}{\line(1,0){0.49}}
\multiput(85.73,26.45)(0.49,0.08){1}{\line(1,0){0.49}}
\multiput(86.22,26.52)(0.49,0.09){1}{\line(1,0){0.49}}
\multiput(86.71,26.61)(0.49,0.1){1}{\line(1,0){0.49}}
\multiput(87.2,26.71)(0.49,0.11){1}{\line(1,0){0.49}}
\multiput(87.69,26.82)(0.48,0.12){1}{\line(1,0){0.48}}
\multiput(88.17,26.94)(0.48,0.13){1}{\line(1,0){0.48}}
\multiput(88.65,27.07)(0.48,0.14){1}{\line(1,0){0.48}}
\multiput(89.13,27.21)(0.47,0.15){1}{\line(1,0){0.47}}
\multiput(89.61,27.36)(0.47,0.16){1}{\line(1,0){0.47}}
\multiput(90.08,27.53)(0.47,0.17){1}{\line(1,0){0.47}}
\multiput(90.54,27.7)(0.23,0.09){2}{\line(1,0){0.23}}
\multiput(91.01,27.88)(0.23,0.1){2}{\line(1,0){0.23}}
\multiput(91.47,28.08)(0.23,0.1){2}{\line(1,0){0.23}}
\multiput(91.92,28.28)(0.23,0.11){2}{\line(1,0){0.23}}
\multiput(92.37,28.49)(0.22,0.11){2}{\line(1,0){0.22}}
\multiput(92.82,28.72)(0.22,0.12){2}{\line(1,0){0.22}}
\multiput(93.26,28.95)(0.22,0.12){2}{\line(1,0){0.22}}
\multiput(93.7,29.19)(0.21,0.13){2}{\line(1,0){0.21}}
\multiput(94.13,29.45)(0.21,0.13){2}{\line(1,0){0.21}}
\multiput(94.55,29.71)(0.21,0.14){2}{\line(1,0){0.21}}
\multiput(94.97,29.98)(0.21,0.14){2}{\line(1,0){0.21}}
\multiput(95.38,30.26)(0.2,0.14){2}{\line(1,0){0.2}}
\multiput(95.79,30.55)(0.2,0.15){2}{\line(1,0){0.2}}
\multiput(96.19,30.85)(0.13,0.1){3}{\line(1,0){0.13}}
\multiput(96.58,31.15)(0.13,0.11){3}{\line(1,0){0.13}}
\multiput(96.96,31.47)(0.13,0.11){3}{\line(1,0){0.13}}
\multiput(97.34,31.79)(0.12,0.11){3}{\line(1,0){0.12}}
\multiput(97.71,32.13)(0.12,0.11){3}{\line(1,0){0.12}}
\multiput(98.08,32.47)(0.12,0.12){3}{\line(1,0){0.12}}
\multiput(98.43,32.82)(0.12,0.12){3}{\line(0,1){0.12}}
\multiput(98.78,33.17)(0.11,0.12){3}{\line(0,1){0.12}}
\multiput(99.12,33.54)(0.11,0.12){3}{\line(0,1){0.12}}
\multiput(99.46,33.91)(0.11,0.13){3}{\line(0,1){0.13}}
\multiput(99.78,34.29)(0.11,0.13){3}{\line(0,1){0.13}}
\multiput(100.1,34.67)(0.1,0.13){3}{\line(0,1){0.13}}
\multiput(100.4,35.06)(0.15,0.2){2}{\line(0,1){0.2}}
\multiput(100.7,35.46)(0.14,0.2){2}{\line(0,1){0.2}}
\multiput(100.99,35.87)(0.14,0.21){2}{\line(0,1){0.21}}
\multiput(101.27,36.28)(0.14,0.21){2}{\line(0,1){0.21}}
\multiput(101.54,36.7)(0.13,0.21){2}{\line(0,1){0.21}}
\multiput(101.8,37.12)(0.13,0.21){2}{\line(0,1){0.21}}
\multiput(102.06,37.55)(0.12,0.22){2}{\line(0,1){0.22}}
\multiput(102.3,37.99)(0.12,0.22){2}{\line(0,1){0.22}}
\multiput(102.53,38.43)(0.11,0.22){2}{\line(0,1){0.22}}
\multiput(102.76,38.88)(0.11,0.23){2}{\line(0,1){0.23}}
\multiput(102.97,39.33)(0.1,0.23){2}{\line(0,1){0.23}}
\multiput(103.17,39.78)(0.1,0.23){2}{\line(0,1){0.23}}
\multiput(103.37,40.24)(0.09,0.23){2}{\line(0,1){0.23}}
\multiput(103.55,40.71)(0.17,0.47){1}{\line(0,1){0.47}}
\multiput(103.72,41.17)(0.16,0.47){1}{\line(0,1){0.47}}
\multiput(103.89,41.64)(0.15,0.47){1}{\line(0,1){0.47}}
\multiput(104.04,42.12)(0.14,0.48){1}{\line(0,1){0.48}}
\multiput(104.18,42.6)(0.13,0.48){1}{\line(0,1){0.48}}
\multiput(104.31,43.08)(0.12,0.48){1}{\line(0,1){0.48}}
\multiput(104.43,43.56)(0.11,0.49){1}{\line(0,1){0.49}}
\multiput(104.54,44.05)(0.1,0.49){1}{\line(0,1){0.49}}
\multiput(104.64,44.54)(0.09,0.49){1}{\line(0,1){0.49}}
\multiput(104.73,45.03)(0.08,0.49){1}{\line(0,1){0.49}}
\multiput(104.8,45.52)(0.07,0.49){1}{\line(0,1){0.49}}
\multiput(104.87,46.01)(0.06,0.5){1}{\line(0,1){0.5}}
\multiput(104.92,46.51)(0.04,0.5){1}{\line(0,1){0.5}}
\multiput(104.97,47.01)(0.03,0.5){1}{\line(0,1){0.5}}
\multiput(105,47.5)(0.02,0.5){1}{\line(0,1){0.5}}
\multiput(105.02,48)(0.01,0.5){1}{\line(0,1){0.5}}

\linethickness{1mm}
\put(30,50){\line(1,0){30}}
\linethickness{1mm}
\multiput(60,90)(0.12,-0.16){125}{\line(0,-1){0.16}}
\linethickness{1mm}
\multiput(104,55)(0.36,0.12){83}{\line(1,0){0.36}}
\linethickness{0.3mm}
\multiput(95,30)(0.12,-0.16){125}{\line(0,-1){0.16}}
\put(35,55){\makebox(0,0)[cc]{$p_1$}}

\put(72,85){\makebox(0,0)[cc]{$p_2$}}

\put(95,80){\makebox(0,0)[cc]{$\cdot$}}

\put(110,70){\makebox(0,0)[cc]{$\cdot$}}

\put(120,55){\makebox(0,0)[cc]{$p_N$}}

\put(110,20){\makebox(0,0)[cc]{$k$}}

\put(80,50){\makebox(0,0)[cc]{$\wt\Gamma$}}

\end{picture}

}

\begin{document}

\baselineskip 24pt

\begin{center}
{\Large \bf  Soft Theorems in Superstring  Theory}

\end{center}

\vskip .6cm
\medskip

\vspace*{4.0ex}

\baselineskip=18pt

\centerline{\large \rm Ashoke Sen}

\vspace*{4.0ex}

\centerline{\large \it Harish-Chandra Research Institute, HBNI}
\centerline{\large \it  Chhatnag Road, Jhusi,
Allahabad 211019, India}



\vspace*{1.0ex}
\centerline{\small E-mail:  sen@mri.ernet.in}

\vspace*{5.0ex}

\centerline{\bf Abstract} \bigskip

We use insights from superstring field theory to prove the subleading
soft graviton theorem for tree amplitudes of
(compactified) heterotic and type II string theories
for arbitrary number of finite energy NS (NSNS) sector external
states but only one soft
graviton. We also prove the leading soft graviton theorem in these theories
for arbitrary number of external soft gravitons. 
In our analysis there is no restriction on the mass and spin
of the finite energy external states. This method can also be used to give a
proof of the subleading soft graviton theorem for tree amplitudes in quantum field
theories coupled to gravity with generic interactions. We discuss the technical issue
involved in extending this analysis to loop amplitudes of superstring theory including
Ramond sector external states, and its possible resolution.

\bigskip



\vfill \eject

\baselineskip 18pt

\tableofcontents

\sectiono{Introduction} \label{sintro}

Study of soft graviton theorem has a long history, both in 
quantum field 
theories\cite{weinberg1,weinberg2,jackiw1,jackiw2,1103.2981,1404.4091,1404.7749,
1405.1015,1405.1410,1405.2346,1405.3413,1405.3533,
1406.6574,1406.6987,1406.7184,
1407.5936,1407.5982,1408.4179,1410.6406,1412.3699,
1503.04816,1504.01364,1507.08882,
1509.07840,1604.00650,1604.03893,
1607.02700,1611.02172,1611.03137,1611.07534,
1702.02350}
and in (super)string theories\cite{ademollo,shapiro,1406.4172,1406.5155,1411.6661,
1502.05258,1505.05854,1506.05789,1507.08829,1511.04921,1601.03457,
1604.03355,1610.03481}. The recent interest in soft graviton
theorem has its origin in the connection
between soft theorems and BMS 
symmetry\cite{1312.2229,1401.7026,1411.5745,1509.01406,1605.09094,1612.08294,1701.00496,1608.00685,1612.05886}.

Our goal in this paper will be to prove certain soft graviton theorems for the
tree amplitudes in heterotic
and type II string theories -- collectively called superstring theory -- possibly
compactified on certain manifolds. We shall use the language of 
superstring field theory\cite{1703.06410}
in which the amplitudes of superstring theory are given as sum of Feynman diagrams
just as in ordinary quantum field theories. However we shall not use many details
of the theory, and for this reason our analysis will apply also to quantum field theories.

The general strategy for computing an amplitude with soft insertions will be as
follows. First we need to identify the Feynman diagrams that give the desired 
contribution. Then in order to evaluate these Feynman diagrams we have to
find the interaction vertices that couple a soft graviton to the rest of the fields.
Once the interaction vertices are found we then use them to evaluate the relevant
diagrams.
The main simplification in our analysis will be in the second step. We follow the
following procedure for determining the coupling of a soft graviton to the rest of
the fields.
\begin{enumerate}
\item Let $\mu,\nu$ denote the coordinate index along the flat
non-compact directions. 
We take the metric $g_{\mu\nu}$ to be sum of three parts: the background
$\eta_{\mu\nu}$, the finite energy part $2\, h_{\mu\nu}$ and the soft part
$2\, S_{\mu\nu}$.
\item In our analysis we treat the finite energy and the soft parts of 
the metric differently.
This is certainly possible for tree amplitudes since soft gravitons appear only as
external particles and in any Feynman diagram there is a clear distinction between
which lines are soft and which lines carry finite energy.\footnote{For 
loop amplitudes we can first
compute the gauge invariant one particle irreducible (1PI) 
effective action without this decomposition into 
soft and finite energy 
parts and then, while computing the full Green's functions by summing
over tree amplitudes using the 1PI effective action, use this decomposition. As long
as the number of non-compact dimensions is sufficiently high so that infrared
and collinear divergences are absent, this is a well defined procedure. However
there are other difficulties with loop amplitudes as discussed in \S\ref{sgen},
and for this reason we postpone discussion of loop amplitudes to a future
publication.}
\item We first set $S_{\mu\nu}$ to zero and expand the action in powers
of $h_{\mu\nu}$ and other fields around the vacuum solution. The resulting action
has manifest Lorentz invariance but not manifest general coordinate invariance.
\item We gauge fix this action by using a Lorentz covariant gauge fixing condition.
\item In the resulting action we now replace $\eta_{\mu\nu}$ by 
$\eta_{\mu\nu}+2\, S_{\mu\nu}$, and all derivatives by covariant derivatives 
computed using the Christoffel connection of the metric 
$\eta_{\mu\nu}+2\, S_{\mu\nu}$. By expanding the resulting action to first order
in $S_{\mu\nu}$ we determine the coupling of the soft graviton to the rest of the
fields.\footnote{Since we shall apply this procedure on the gauge fixed action, 
this makes the gauge fixing terms for the finite energy fields
covariantized with respect to the soft graviton.}
\item This generates the coupling of the
soft graviton to the rest of the fields, including finite energy components of the
metric, up to first order in the derivatives but misses the terms involving two or
more derivatives of $S_{\mu\nu}$ from possible couplings via the Riemann tensor
computed from $\eta_{\mu\nu}+2\, S_{\mu\nu}$. Therefore our results are valid to first 
subleading order
in the momentum of the soft particle.
\item This action is invariant under general coordinate transformation of $S_{\mu\nu}$.
But since $S_{\mu\nu}$ will only appear as external line, we do not need to fix any
gauge for $S_{\mu\nu}$. Indeed we only make use of part of this action containing 
terms linear in $S_{\mu\nu}$ and not the full action.
\item We can now replace $S_{\mu\nu}$ by $\ve_{\mu\nu} e^{ik.x}$ to determine the
coupling of a soft graviton of momentum $k$ and polarization $\ve$. Since the 
coupling is determined by replacing $\eta_{\mu\nu}$ by $\eta_{\mu\nu}+2\,
S_{\mu\nu}$,
it makes computing the effect of soft graviton coupling simple. For example if
we consider part of an amplitude that has a finite limit when the momentum of
the soft graviton goes to zero, and if we want to compute just the leading term
of this component
in the power series expansion in $k$, we simply have to study the effect of
replacing $\eta_{\mu\nu}$ by $\eta_{\mu\nu}+2\, \ve_{\mu\nu}$. Instead of 
studying its effect on each vertex and internal propagator, we can determine the
result by making this replacement in the final expression for the original amplitude
without the soft graviton.
\end{enumerate}

Using this method, we prove that at tree level
the subleading soft graviton theorem given in 
\cite{1312.2229,1401.7026,1404.4091}
holds for one soft graviton and arbitrary number of finite energy
external
states coming from the NS sector in the heterotic string theory and the NSNS
sector in the type II string theory. 
There is no restriction on the mass and spin of the finite energy external
states in either analysis. We also generalize the leading order  
result to the case where there
are multiple soft gravitons. We believe that
it should be possible to use the method suggested here to prove
the subleading soft 
graviton theorem
for string loop amplitudes and Ramond sector external states
as well. We discuss in \S\ref{sgen} the
main technical difficulty in proving this general result. 
However, since
this method misses terms involving Riemann tensor of the soft graviton, which 
contains two powers of the soft momentum, it cannot be 
used to express the subsubleading soft graviton amplitude just in terms of the
amplitude without the soft graviton. We must separately take into account the effect
of the extra terms proportional to the Riemann tensor of the soft graviton, and
such terms vary from one theory to another.
This is consistent with the fact
that subsubleading soft graviton amplitudes are known to be
non-universal\cite{1604.03355} 
due to non-minimal coupling of the metric via the Riemann tensor.

Our method can also be generalized to derive the soft photon 
theorem\cite{low} using the
same principle: the coupling of a soft photon to the rest of the fields is determined
by making all derivatives into covariant derivatives using the soft photon field.
In this case only the leading soft photon theorem is universal since subleading
coupling of soft photons can be modified via non-minimal coupling involving
field strength. 

The rest of the paper is organized as follows. In \S\ref{swarm} we consider a
scalar field theory coupled to gravity in arbitrary dimensions with arbitrary
interactions, and show how in this theory 
we can derive the subleading soft graviton theorem for tree level
amplitudes with one external soft graviton and arbitrary number of finite energy 
external scalars. In \S\ref{stree}
we generalize this method to prove subleading soft graviton theorem for 
tree amplitudes in superstring
theory with one external soft graviton and arbitrary number of finite energy
external states from the NS sector. In \S\ref{smulti} we extend the result
of \S\ref{stree} to the case where there are multiple soft gravitons, but work only
to the leading order in the soft momentum.
In \S\ref{sgen}
we extend the results
to leading soft photon theorem, and also
discuss the possible ways of extending the results to loop amplitudes in
superstring theory.

\sectiono{Warm up with scalar field coupled to gravity} \label{swarm}

In this section we shall consider a theory of a scalar field $\phi$ coupled to
gravity in $D$ space-time dimensions, providing a proof of soft graviton theorem
that is slightly different from the one given {\it e.g.} in 
\cite{1406.6574,1406.6987}.
The scalar field can be massive or massless.
Furthermore, we shall not put any restriction on the
interactions, except the requirement of invariance under general coordinate 
transformation. 
We shall use the convention that an external or 
internal massless particle in a Feynman diagram will be called soft if all the 
components of its momentum are small in the center of momentum frame. 
On the other hand if an internal particle carries a momentum $p$ that has one 
or more components large but 
$p^2 +M^2$, where $M$ denotes the mass of the particle, is small then it will be 
called a nearly on-shell particle. Finally if $p^2 +M^2$ is of order unity or larger, 
then it will be called a hard particle.

\begin{figure}
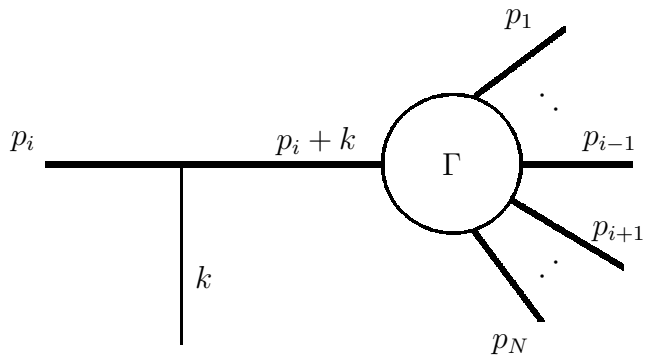


\begin{center}

\figsoftonefieldtree

\end{center}

\caption{Source of the leading contribution to the amplitude with one external
soft graviton. 
\label{f1fieldtree}
}

\end{figure}

\begin{figure}
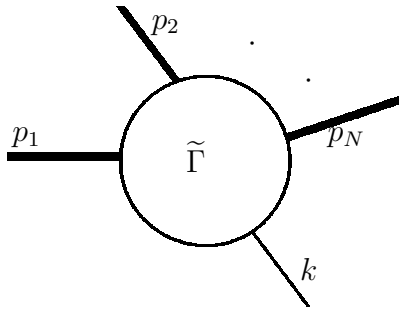


\begin{center}

\figsoftthreefieldtree

\end{center}

\caption{Source of the subleading contribution to the amplitude with one external
soft graviton. Here $\wt\Gamma$ denotes tree amplitudes with external
propagators removed, and also diagrams of the type given in Fig.~\ref{f1fieldtree}, where
the soft graviton attaches to one of the external finite energy lines, removed.
\label{f3fieldtree}
}

\end{figure}

Let us now consider a tree amplitude where one external soft graviton 
carrying momentum $k$ is attached to a graph with $N$ external on-shell scalar
particles carrying finite  momenta $p_1,\ldots , p_N$.
All external momenta will be taken to be ingoing in the Feynman diagram, i.e.\
outgoing particles will carry negative $p^0$. Our goal will
be to compute the $k$ dependence of the amplitude to order unity. 
The leading contribution to this amplitude, of order $k^{-1}$, comes from the
diagram shown in Fig.~\ref{f1fieldtree}. In this figure
the thick lines denote either external on-shell finite energy particles
or internal nearly on-shell particles, whereas the thin line denotes the external
soft particle. $\Gamma$ denotes amputated tree level
Green's function -- Green's functions
from which the external propagators are removed. If $M$
denotes the  mass of the scalar field, then for small $k$ the internal
propagator carrying momentum $p_i+k$  can generate a large factor proportional to
\be \label{enosh1}
\{(p_i+k)^2 +M^2\}^{-1} = (2p_i\cdot k)^{-1}\, .
\ee
It is easy to see that the diagrams given in Fig.~\ref{f1fieldtree}
are the only ones that can produce a soft factor in the denominator.
At the first subleading order, i.e.\ at order unity, the contribution to the amplitude
comes from the subleading contributions from Fig.~\ref{f1fieldtree}, as well as
the leading contribution from the diagram shown in Fig.~\ref{f3fieldtree}. 

In order to compute soft graviton amplitudes, we need the know the coupling of a 
soft graviton to the rest of the fields. For this we follow the strategy outlined in
\S\ref{sintro}:
\begin{enumerate}
\item First we write the metric as $\eta_{\mu\nu}+2\, h_{\mu\nu}$ and expand the action
in a power series in $h_{\mu\nu}$ and $\phi$. We shall use the Feynman rules derived
from this action to compute
amplitudes involving finite energy external particles but not external soft particles. For
the analysis of this section we shall take the external states to be $\phi$ 
particles. \label{is1}
\item In order to compute the amplitude involving a soft graviton of momentum 
$k$ and polarization
$\ve_{\mu\nu}$ satisfying\footnote{Throughout our discussion all indices
will be raised and lowered by $\eta^{\mu\nu}$ and $\eta_{\mu\nu}$.}
\be \label{egcond}
\eta_{\mu\nu} \ve^{\mu\nu}=0, \quad  \ve^{\mu\nu} = \ve^{\nu\mu}, \quad
k_\mu \ve^{\mu\nu} = 0, \quad k_\nu \ve^{\mu\nu}=0\, ,
\ee 
we replace, in the action computed in step \ref{is1},
$\eta_{\mu\nu}$ by $\eta_{\mu\nu}
+ 2\, \ve_{\mu\nu} e^{ik.x}$ and all ordinary derivatives by covariant derivatives
computed with this metric and expand the action to first order in $\ve_{\mu\nu}$. 
Since we have taken the polarization tensor to be traceless, we do not need
to worry about the $\sqrt{\det g}$ term in the action. This
determines the coupling of the soft graviton to the rest of the fields (scalars and
finite energy gravitons) correctly to linear order in momentum $k^\mu$.
There may be additional coupling at quadratic and higher order in $k^\mu$ through
couplings involving Riemann tensor that are missed by this expansion. 
\end{enumerate}

We shall now proceed to evaluate the contribution from Fig.~\ref{f1fieldtree}. 
Using the prescription given above, the coupling of the soft graviton to the
scalar field is given by replacing $\eta^{\mu\nu}$ by 
$\eta^{\mu\nu}-2\, \ve^{\mu\nu} e^{ik.x}$ in the quadratic term in the action
involving the scalar field:
\be \label{escalarproptree}
-{1\over 2}
\int d^D x \{ \eta^{\mu\nu} \p_\mu \phi \p_\nu \phi + M^2 \phi^2\} \, .
\ee
We have assumed that the quadratic term contains only two derivatives. If
it contained higher powers of $(-\square + M^2)$, 
they could be removed by a redefinition
of $\phi$ that is regular near on-shell field configuration, and therefore would not
affect the S-matrix.
This gives the following coupling of the scalar to the soft graviton:
\be
\int d^D x \, \ve^{\mu\nu} e^{ik.x} \p_\mu \phi \p_\nu \phi \, .
\ee
Therefore the contribution from the three point vertex in Fig.~\ref{f1fieldtree}
involving the soft graviton and the external scalar is given by
\be \label{eficontree}
 2\, i \, \ve^{\mu\nu}  p_{i\, \mu} (p_i+k)_{\nu}
= 2\, {i} \, \ve^{\mu\nu}  p_{i\, \mu} p_{i\, \nu}\, ,
\ee
where in the last step we have used \refb{egcond}.
The second piece in Fig.~\ref{f1fieldtree}
is the contribution from the scalar propagator carrying momentum $(p_i+k)$.
This is given by
\be \label{esecontree}
-i \, {1\over (p_i+k)^2  + M^2} = -i \, {1\over 2 p_i\cdot k}\, ,
\ee
using the on-shell conditions $p_i^2+M^2=0$, $k^2=0$.
The final piece in  Fig.~\ref{f1fieldtree}  is the 
contribution from the amputated Green's
function
with external scalar particles of momenta $p_1,\ldots , p_{i-1},
p_i+k, p_{i+1}, \ldots , p_N$. This is given by
\be \label{ethcontree}
\Gamma(p_1, \ldots , p_N) + k_\rho {\p\over \p p_{i\, \rho}} 
\Gamma(p_1,\ldots , p_N)
+\OO(k_\mu k_\nu)\, .
\ee
Taking the product of \refb{eficontree}, \refb{esecontree} and 
\refb{ethcontree} we get the full
contribution from Fig.~\ref{f1fieldtree} to order unity. We can now sum over all
diagrams in which the soft graviton is attached to any of the external states $i$.
The net contribution from all these diagrams is given by
\be \label{efullone}
\sum_{i=1}^N 
\ve^{\mu\nu} p_{i\,\mu} p_{i\,\nu} \, {1\over  p_i\cdot k} \, \Gamma(p_1, \ldots , p_N) 
+ \sum_{i=1}^N \ve^{\mu\nu} p_{i\,\mu} p_{i\,\nu}  \, {1\over  p_i\cdot k} \,  
k_\rho {\p\over \p p_{i\, \rho}} \Gamma(p_1,\ldots , p_N) + \OO(k)\, .
\ee

The first term gives the amplitude to order $1/k$, while the second term
is of order unity. However if we want to compute
the total order unity contribution, we also need to compute the contribution from
the diagrams shown in Fig.~\ref{f3fieldtree}. In this figure $\wt\Gamma$ 
denotes the
sum of Feynman diagrams from which external propagators have been removed,
and also diagrams like the ones shown in Fig.~\ref{f1fieldtree}, in which by
cutting a single internal propagator we can remove the soft line and one more
external line from the rest of the diagram, have been removed. Therefore in
Fig.~\ref{f3fieldtree} the soft
graviton is attached to a hard internal line, and the amplitude has a 
finite limit as $k^\mu \to 0$. This in turn means that to evaluate these diagrams
to order unity, we can set the momentum of the external soft graviton to zero.
Using the rules for determining the soft graviton interaction vertex described earlier,
we see that
such an amplitude has the interpretation of a deformation of the amputated
Green's function without the soft graviton under a constant change in the background
metric $\eta^{\mu\nu}\to \eta^{\mu\nu}-2\, \ve^{\mu\nu}$ to first order in $\ve_{\mu\nu}$.
Since $\wt\Gamma(p_1,\ldots p_N)$ depends on the metric only via the combinations
$g^{\mu\nu} p_{i\, \mu} p_{j\, \nu} = (\eta^{\mu\nu} -2\, 
\ve^{\mu\nu}) p_{i\, \mu} p_{j\, 
\nu}$,
the effect of deforming the background
metric by $2\, \ve_{\mu\nu}$ can also be equivalently
represented by deforming $p_{i\, \mu}$ to $p_{i\, \mu} - 
\ve_\mu^{~\sigma} p_{i\, \sigma}$. Therefore
the contribution from Fig.~\ref{f3fieldtree} can be expressed as
\be \label{efulltwo}
-\sum_{i=1}^N \,  \ve_\mu^{~\nu} \, p_{i\, \nu} \, {\p\over \p p_{i\, \mu}}
\Gamma(p_1,\ldots p_N)\, .
\ee

Adding \refb{efullone} and \refb{efulltwo} we get the full amplitude to order
unity:
\ben \label{efullsofttree}
 &&  \sum_{i=1}^N 
\ve^{\mu\nu} p_{i\, \mu} p_{i\, \nu} \, {1\over p_i\cdot k} \, \Gamma
(p_1, \ldots , p_N) 
\nonumber \\
&& +  \sum_{i=1}^N \left[\ve^{\mu\nu} p_{i\, \mu} p_{i\, \nu} \, {1\over p_i\cdot k} \,  
k^\rho - \ve_\rho^{~\nu} \, p_{i\, \nu}\right]  
{\p\over \p p_{i\, \rho}} \Gamma(p_1,\ldots , p_N) + \OO(k)\, .
\een
This is the subleading soft graviton theorem
for one external soft
graviton\cite{1312.2229,1401.7026,1404.4091}.

Note that even if we set the external states on-shell by setting $p_i^2+M^2=0$,
computation of $\p \Gamma /\p p_{i\,\rho}$ requires off-shell
information. For example if we add to $\Gamma$ a contribution proportional to
$(p_i^2+M^2)$ that vanishes on-shell, $\p \Gamma /\p p_{i\, \rho}$ gets a
contribution proportional to $p_{i}^{\rho}$ that does not vanish on-shell.
However when substituted into \refb{efullsofttree} it does vanish, showing that
\refb{efullsofttree} depends only on on-shell data.

\sectiono{One soft graviton theorem in tree level superstring field theory} \label{stree}

We now turn to superstring field theory\cite{1703.06410}. It can be regarded as
a regular field theory
of infinite number of fields of arbitrarily high spin, 
with interaction vertices that are exponentially suppressed at large
euclidean momenta. This makes the contribution from each Feynman graph manifestly
ultraviolet finite, but otherwise the amplitudes are expressed as sum over Feynman
diagrams just as an ordinary quantum field theory. 
Therefore as in \S\ref{swarm}, the leading contribution to the soft graviton amplitude
will come from Fig.~\ref{f1fieldtree}, and the subleading contribution will come from
Fig.~\ref{f1fieldtree}  and Fig.~\ref{f3fieldtree}, although the Feynman rules for
evaluating these diagrams will be different.
We shall now evaluate these contributions by 
restricting the external
states to be from the
NS sector in the heterotic string theory and NSNS sector in type II string
theory. During this analysis 
we shall allow for the possibility that some of the spatial directions have
been compactified, and denote by $D$ the number of non-compact space-time
dimensions. 

Note that even though we make use of superstring field theory, at tree level amplitudes
computed from this theory are identical to the standard amplitudes of superstring theory
computed using world-sheet methods. Therefore our proof of subleading soft graviton 
theorem holds for the standard amplitudes computed using the world-sheet methods.
At loop level the world-sheet approach gives divergent results when the masses of the
external states are renormalized, but superstring field theory continues to give sensible
S-matrix elements via the standard LSZ framework.

Before proceeding to the details of the analysis,
let us comment on one underlying assumption that will be made in
our analysis. In superstring field theory the graviton is a specific component of
the string field. Therefore coupling of a soft graviton to the rest of the fields will be
determined by the change in the interaction vertices / propagators due to the
effect of  switching on a low momentum
plane wave solution of this field to first order in the field.
We shall be using the fact that to linear order, this deformation 
is equivalent to a deformation of the world-sheet superconformal
field theory underlying the construction of the superstring field theory 
due to a change in
the
target space metric $\eta_{\mu\nu}$ to $\eta_{\mu\nu}+2\,
\ve_{\mu\nu}e^{ik.x}$.  
In the string
field theory literature this property is known as background independence. Background
independence of closed bosonic string field theory was established in 
\cite{9307088,9311009}. 
This has not yet been proved for superstring field theory, but there
does not seem to be any specific difficulty in doing so\cite{prepare}. Our
analysis will assume background independence of superstring field theory, since
we shall be computing the coupling of soft graviton by studying the effect of deforming
the target space metric entering the construction of the world-sheet superconformal
field theory.

Let us suppose that the $i$-th external particle is associated with some rank
$n_i$ tensor field $\phi_{\mu_1\ldots \mu_{n_i}}$. In the Siegel gauge the kinetic
operator acting on the NS sector states is proportional to $(L_0+\bar L_0)$ where
$L_n,\bar L_n$ are the world-sheet Virasoro generators. Acting on a state of
momentum $p$ and mass $M$ this is proportional
to $p^2+M^2$.
Therefore we can choose
a basis for the NS sector string fields in which the kinetic
term of $\phi_{\mu_1\ldots \mu_{n_i}}$ takes a particularly
simple form\footnote{Note that we have taken the field $\phi$ to be  a covariant tensor.
While coupling it to a background soft graviton we shall take these to be 
the independent
fields. This is of course related by field redefinition to the case where the fields with
contravariant indices are regarded as independent fields. The S-matrix is independent
of which prescription we choose.}
\be \label{eactionsft}
-{1\over 2} \int d^D x   \prod_{j=1}^{n_i} \eta^{\mu_j\nu_j} \left[
\eta^{\mu \nu}
\p_\mu \phi_{\mu_1\ldots \mu_{n_i}} 
\p_\nu \phi_{\nu_1\ldots \nu_{n_i}} 
+ M_i^2 \, \phi_{\mu_1\ldots \mu_{n_i}} \phi_{\nu_1\ldots \nu_{n_i}} \right]\, ,
\ee
where $M_i$ denotes the tree level mass of the field $\phi_{\mu_1\ldots \mu_{n_i}}$.
Typically the tensor has specific symmetry properties, but we can
ignore this for now and restore it at the end by choosing the polarizations
of the external states to have the required symmetry. Eq.\refb{eactionsft} is the 
only specific property of the superstring field theory action that 
we shall use -- we shall not need to use any detailed property of the interaction
terms except general coordinate invariance of the action. From
\refb{eactionsft} we see that
the coupling of the soft graviton to the tensor field $\phi$ has three
types of terms:
\begin{enumerate}
\item The first type of term is given by the replacement of $\eta^{\mu\nu}$ by
$\eta^{\mu\nu}-2\, \ve^{\mu\nu} e^{ik.x}$ in \refb{eactionsft}. 
The effect of this is identical to that given in 
\refb{eficontree}, and generates a contribution to the vertex:
\be \label{essone1}
2\, {i} \, \ve^{\mu\nu} p_{i\, \mu} p_{i\, \nu} \prod_{j=1}^{n_i}\eta^{\mu_j \nu_j}\, , 
\ee
where $\mu_1,\ldots , \mu_{n_i}$ are the Lorentz indices carried by the external state
carrying momentum $p_i$ and $\nu_1,\ldots , \nu_{n_i}$ are the Lorentz indices
carried by the internal state carrying momentum $p_i+k$.
\item
The second kind of contribution comes from replacing in \refb{eactionsft} the
$\eta^{\mu_\ell\nu_\ell}$ factor by $\eta^{\mu_\ell\nu_\ell} - 2\,
\ve^{\mu_\ell\nu_\ell} e^{ik.x}$ for $1\le \ell \le n_i$.
The effect of this is to generate an interaction vertex
\be \label{essone2}
2\, {i} \, \{\eta^{\mu\nu} p_{i\, \mu} (p_i+k)_{\nu} + M_i^2\}
\sum_{\ell=1}^{n_i} \ve^{\mu_\ell \nu_\ell} \, 
\prod_{j=1\atop j\ne \ell}^{n_i}\eta^{\mu_j \nu_j}
=2\, i \, p_i\cdot k\, \sum_{\ell=1}^{n_i} \ve^{\mu_\ell \nu_\ell} \, 
\prod_{j=1\atop j\ne \ell}^{n_i}\eta^{\mu_j \nu_j}\, ,
\ee
where in the second step we have used the on-shell condition $p_i^2+M_i^2=0$.
\item
The third type of contribution comes from replacing the $\p_{\mu}\phi_{\mu_1\ldots
\mu_{n_i}}$ (or
$\p_{\nu}\phi_{\nu_1\ldots \nu_{n_i}}$) factor by
\ben \label{ecov}
&& D_{\mu}  \phi_{\mu_1\ldots
\mu_{n_i}}
= \p_\mu \phi_{\mu_1\ldots
\mu_{n_i}} - \sum_{\ell=1}^{n_i} \Gamma_{\mu \mu_\ell}^{\rho_\ell}
\phi_{\mu_1\ldots  \mu_{\ell-1} \rho_\ell \mu_{\ell+1}\ldots \mu_{n_i}}
 \\
&=&  \p_\mu \phi_{\mu_1\ldots
\mu_{n_i}} - {i}  \, e^{ik.x} 
\sum_{\ell=1}^{n_i} (k_\mu \ve^{\rho_\ell}_{~\mu_\ell}
+ k_{\mu_\ell} \ve^{\rho_\ell}_{~\mu} - k^{\rho_\ell} \ve_{\mu\mu_\ell}) 
\phi_{\mu_1\ldots  \mu_{\ell-1} \rho_\ell \mu_{\ell+1}\ldots \mu_{n_i}} +
\OO(k_\rho k_\sigma)\, . \nonumber
\een
This, together with similar expression for $D_{\nu}  \phi_{\nu_1\ldots
\nu_{n_i}}$, generates the following net 
contribution to the vertex to first order in the soft
momenta:
\ben \label{essone3}
&& {i}\, \eta^{\mu\nu} (p_i +k)_\nu \sum_{\ell=1}^{n_i}
(k_\mu \ve^{\mu_\ell\nu_\ell}
+ k^{\nu_\ell} \ve^{\mu_\ell}_{~\mu} - k^{\mu_\ell} \ve_{\mu}^{~\nu_\ell})
 \prod_{j=1\atop j\ne \ell}^{n_i}
\eta^{\mu_j\nu_j} \nonumber \\
&& - {i} \, \eta^{\mu\nu} p_{i\,\mu} \sum_{\ell=1}^{n_i}
(k_\nu \ve^{\nu_\ell\mu_\ell}
+ k^{\mu_\ell} \ve^{\nu_\ell}_{~\nu} - k^{\nu_\ell} \ve_{\nu}^{~\mu_\ell})
 \prod_{j=1\atop j\ne \ell}^{n_i}
\eta^{\mu_j\nu_j} \, .
\een
Here $\mu_1,\ldots\, \mu_{n_i}$ are the Lorentz indices carried by the external
line of momentum $p_i$ and $\nu_1,\ldots , \nu_i$ are the Lorentz indices carried
by the internal line of momentum $p_i+k$.\end{enumerate}
Adding \refb{essone1}, \refb{essone2} and \refb{essone3} and keeping terms
up to order $k$ we get the following net contribution to the three point vertex
of Fig.~\ref{f1fieldtree} with
one soft graviton carrying momentum $k$ and a pair of $\phi$ particles
carrying momenta $p_i$ and $-p_i-k$: 
\ben \label{essone}
2\, {i} \, \ve^{\mu\nu} p_{i\, \mu} p_{i\, \nu} \prod_{j=1}^{n_i}\eta^{\mu_j \nu_j}
+2\, i \sum_{\ell=1}^{n_i} 
\left[ p_i\cdot k \, \ve^{\mu_\ell \nu_\ell} 
- p_{i\, \mu} \ve^{\mu \nu_\ell} k^{\mu_\ell} + p_{i\, \mu} \ve^{\mu \mu_\ell} k^{\nu_\ell}  
\right]  \prod_{j=1\atop j\ne \ell}^{n_i} 
\eta^{\mu_j\nu_j} +\OO(k_\rho k_\sigma)\, .
\een

The next task is to compute the $\phi$ propagator carrying momentum $p_i+k$.
This is easily read from \refb{eactionsft} to be
\be \label{esstwo}
-i \, \prod_{j=1}^{n_i} \eta_{\nu_j \rho_j} \, {1\over (p_i+k)^2 + M^2}
= -i \, \prod_{j=1}^{n_i} \eta_{\nu_j \rho_j} \, {1\over 2 p_i\cdot k}\, ,
\ee
where $\rho_1,\ldots , \rho_{n_i}$ are the tensor indices carried by the right end
of the internal propagator carrying momentum $p_i+k$.

Finally we turn to the part of the amplitude $\Gamma$ with external
states carrying momenta $p_1,\ldots , p_{i-1}, p_i+k, p_{i+1},\ldots , p_N$.
If $\rho_1\ldots , \rho_{n_i}$ are the tensor indices carried by the leg of
momentum $p_i+k$ entering this amplitude, 
then we shall denote the amplitude by
$\Gamma^{\rho_1\ldots  \rho_{n_i}}$,
suppressing the  tensor indices associated with the other external
states of $\Gamma$.
$\Gamma^{\rho_1\ldots  \rho_{n_i}}$
 may
be expressed as
\ben \label{essthree}
&& \Gamma^{\rho_1\ldots  \rho_{n_i}}(p_1,\ldots , p_{i-1}, p_i+k, p_{i+1},\ldots ,
p_N) \nonumber \\
&=& \Gamma^{\rho_1\ldots  \rho_{n_i}}(p_1,\ldots ,  p_N) 
+ k_\sigma {\p \over \p p_{i\, \sigma} }
\Gamma^{\rho_1\ldots  \rho_{n_i}}(p_1,\ldots ,  p_N) + \OO(k_\mu k_\nu)\, .
\een

The net contribution to Fig.~\ref{f1fieldtree} to order unity is now obtained
by taking the product of \refb{essone}, \refb{esstwo}, \refb{essthree}
and the polarization tensor $\eps_{\mu_1\ldots  \mu_{n_i}}$ of the external
$\phi$ field, expanding
the result to order unity and finally summing over $i$. This gives
\ben \label{efirsttree}
&&\sum_{i=1}^{N}  \ve^{\mu\nu} \, p_{i\, \mu} p_{i\, \nu} \, {1\over  p_i\cdot k}\, 
\eps_{\mu_1\ldots  \mu_{n_i}} \Gamma^{\mu_1\ldots  \mu_{n_i}}
+ \, \sum_{i=1}^N  \ve^{\mu\nu} \, p_{i\, \mu} p_{i\, \nu} \,
{1\over  p_i\cdot k} \, \eps_{\mu_1\ldots  \mu_{n_i}}
k_\sigma {\p \over \p p_{i\, \sigma} } \Gamma^{\mu_1\ldots  \mu_{n_i}}
\nonumber \\
&& + \sum_{i=1}^N \eps_{\mu_1\ldots  \mu_{n_i}} \sum_{\ell=1}^{n_i} 
\, {1\over p_i\cdot k}\, 
\left[ p_i\cdot k \, \ve^{\mu_\ell}_{~\nu_\ell} 
- p_{i\, \mu} \ve^{\mu}_{~\nu_\ell} k^{\mu_\ell} + p_{i\, \mu} \ve^{\mu \mu_\ell} k_{\nu_\ell}  
\right] \Gamma^{\mu_1\ldots  \mu_{\ell-1} \nu_\ell \mu_{\ell+1}\ldots  \mu_{n_i}}\, .
\een

To this we have to add the leading contribution from Fig.~\ref{f3fieldtree}. This
is given by the change in the amplitude $\eps_{\mu_1\ldots  \mu_{n_i}}
\Gamma^{\mu_1\ldots  \mu_{n_i}}$ under the variation $\eta^{\rho\sigma}\to 
\eta^{\rho\sigma}
-2\, \ve^{\rho\sigma}$. Since the final expression for the amplitude is given by
products of $\eps_{\mu_1\ldots  \mu_{n_i}}$'s and $p_{i\, \mu}$'s contracted with 
various factors of $\eta^{\rho\sigma}$,
changing $\eta^{\rho\sigma}$ to $\eta^{\rho\sigma}-2 \, \ve^{\rho\sigma}$ can 
also be implemented
by the change
\be
p_\mu \to p_\mu -  \ve_{\mu}^{~\nu} p_\nu, \qquad
\eps_{\mu_1\ldots  \mu_{n_i}} \to\eps_{\mu_1\ldots  \mu_{n_i}} 
-  \sum_{\ell=1}^{n_i} \ve_{\mu_\ell}^{~\nu_\ell}
\eps_{\mu_1\ldots  \mu_{\ell-1}\nu_\ell \mu_{\ell+1}\ldots  \mu_{n_i}}\, .
\ee
This determines the amplitude given in Fig.~\ref{f3fieldtree} to be
\be \label{esecondtree}
-  \sum_{i=1}^N \ve_{\mu}^{~\nu} \, \eps_{\mu_1\ldots  \mu_{n_i}}  \, 
p_{i\, \nu} {\p \over \p p_{i\, \mu}} 
\Gamma^{\mu_1\ldots  \mu_{n_i}}(p_1,\ldots , p_N)
- \sum_{i=1}^{N}  \sum_{\ell=1}^{n_i} \ve_{\mu_\ell}^{~\nu_\ell}
\eps_{\mu_1\ldots  \mu_{\ell-1}\nu_\ell \mu_{\ell+1}\ldots  \mu_{n_i}} 
\Gamma^{\mu_1\ldots  \mu_{n_i}}\, .
\ee

Adding \refb{efirsttree} to \refb{esecondtree} we now get the full contribution to the
amplitude of one soft graviton and $N$ finite energy particles to the first subleading
order in the soft momentum:
\ben \label{efintree}
&& \left[\sum_{i=1}^{N}  \ve^{\mu\nu} \, p_{i\, \mu} p_{i\, \nu} \, {1\over p_i\cdot k}
\right] \, 
\Gamma(p_1,\ldots , p_N)
\nonumber \\ &&
+ \sum_{i=1}^N
 \eps_{\mu_1\ldots  \mu_{n_i}}
\left( \ve^{\mu\nu} \, p_{i\, \mu} p_{i\, \nu} \, k_\sigma\, 
{1\over  p_i\cdot k}  - \ve_\sigma^{~\nu} p_{i\, \nu}\right) {\p \over \p p_{i\, \sigma} } \Gamma^{\mu_1\ldots  \mu_{n_i}}
\nonumber \\
&& - \sum_{i=1}^N 
{1\over p_i\cdot k} \eps_{\mu_1\ldots  \mu_{n_i}} \sum_{\ell=1}^{n_i} 
\left[  p_{i\, \mu} \ve^{\mu}_{~\nu_\ell} k^{\mu_\ell} - p_{i\, \mu} \ve^{\mu \mu_\ell} k_{\nu_\ell}  
\right] \Gamma^{\mu_1\ldots  \mu_{\ell-1} \nu_\ell \mu_{\ell+1}\ldots  \mu_{n_i}}\, .
\een
In the first term $\Gamma$ denotes the amplitude with the same external finite
energy states
but without soft graviton insertion -- being equal to $\eps_{\mu_1\ldots  \mu_{n_i}} 
\Gamma^{\mu_1\ldots  \mu_{n_i}}$ in the notation of \refb{esecondtree}. In the 
second and the third line it is understood that in the $i$-th term inside the sum, we
have suppressed the polarization tensor of all states other than the $i$-th state and
the corresponding indices of $\Gamma$.

Eq.~\refb{efintree} is the soft graviton theorem to first subleading 
order\cite{1312.2229,1401.7026,1404.4091}.

\sectiono{Multiple soft gravitons} \label{smulti}

\begin{figure}
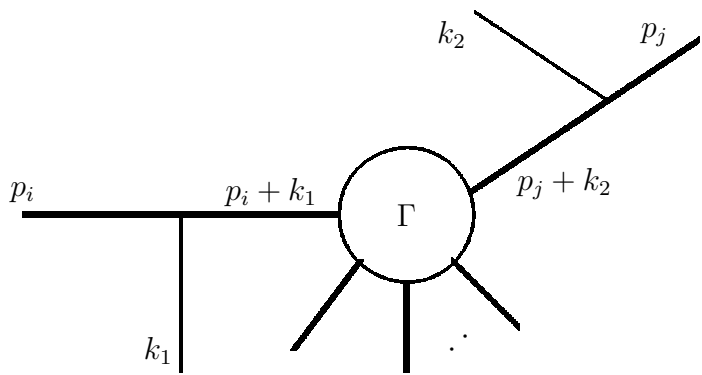


\begin{center}

\figsofttwotree

\end{center}

\caption{Two external soft gravitons attached to different external lines carrying finite momenta.
\label{f2tree}
}

\end{figure}

\begin{figure}
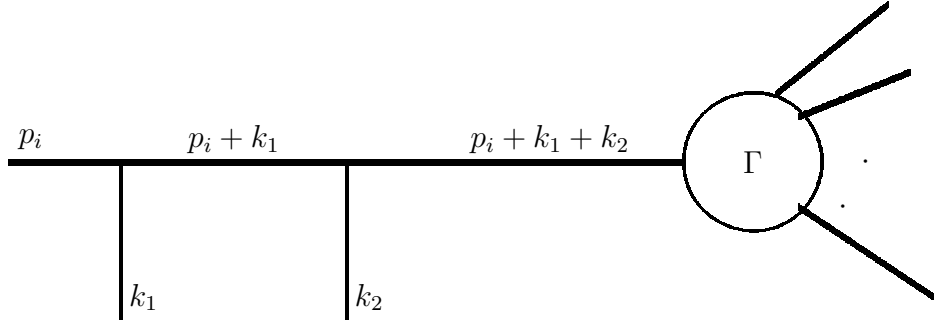


\begin{center}

\figsoftthreetree

\end{center}

\caption{Two external soft gravitons attached to the same external line 
carrying finite momenta.
\label{f3tree}
}

\end{figure}

We shall now consider amplitudes with multiple soft gravitons but restrict our analysis
to the leading order in the soft momenta.  This analysis will be identical to that in
\cite{weinberg2}. 
Consider first the case where we have two soft external gravitons 
carrying momenta $k_1$
and $k_2$. The maximum power of soft momenta in the denominator of such an amplitude
is two. This can arise from the two soft gravitons attaching on different external legs
as in Fig.~\ref{f2tree} or both soft gravitons attaching on the same external leg as in
Fig.~\ref{f3tree}.
In both diagrams, we can compute the product of 
the leading contributions from the three point
vertex from \refb{essone1} and the propagator that follows it 
from \refb{esstwo}.
When the two soft gravitons attach to different external lines as in Fig.~\ref{f2tree},
the amplitude
takes the form
\be 
 {1\over p_i\cdot k_1} \varepsilon^{(1)}_{\mu\nu} p_i^\mu p_i^\nu
 \, \, \times \, \, 
  {1\over p_j\cdot k_2} \varepsilon^{(2)}_{\rho\sigma} p_j^\rho p_j^\sigma
 \, \,  \times \, \,     \Gamma^{(N)}(p_1,\ldots , p_N) \, \,  + \, \,  \hbox{less singular terms}
 \, , \ee
 where $\Gamma^{(N)}(p_1,\ldots , p_N)$ denotes the amplitude without soft gravitons
 with all indices contracted with the external polarization tensors.
On the other hand when both soft lines attach to the same external
line as in Fig.~\ref{f3tree} the amplitude
takes the form
\be 
 {1\over p_i\cdot k_1} \varepsilon^{(1)}_{\mu\nu} p_i^\mu p_i^\nu
 \, \,  \times \, \, 
  {1\over p_i\cdot (k_1+k_2)} \varepsilon^{(2)}_{\rho\sigma} p_j^\rho p_j^\sigma
  \, \,  \times \, \,     \Gamma^{(N)}(p_1,\ldots , p_N) \, \,  
  + \, \,  \hbox{less singular terms}
 \, . \ee
 Adding to this another contribution where the external soft lines carrying momenta
 $k_1$ and $k_2$ are exchanged we get
 \be 
  {1\over p_i\cdot k_1} \varepsilon^{(1)}_{\mu\nu} p_i^\mu p_i^\nu
 \, \,  \times \, \, 
  {1\over p_i\cdot k_2} \varepsilon^{(2)}_{\rho\sigma} p_i^\rho p_i^\sigma
  \, \,  \times \, \,       \Gamma^{(N)}(p_1,\ldots , p_N)  \, \,  + \, \,  \hbox{less singular terms}
 \, . \ee
 Summing over all possible insertions of the two soft lines on $N$ external lines
 carrying finite momentum, we now get\cite{weinberg2}
 \be \label{etwosoftfieldtree}
 \sum_{i=1}^N {1\over p_i\cdot k_1} \varepsilon^{(1)}_{\mu\nu} p_i^\mu p_i^\nu
 \, \,  \times \, \,  \sum_{j=1}^N
  {1\over  p_j\cdot k_2} \varepsilon^{(2)}_{\rho\sigma} p_j^\rho p_j^\sigma
  \, \,  \times \, \,    \Gamma^{(N)}(p_1,\ldots , p_N) 
   \, \,  + \, \,  \hbox{less singular terms}
 \, . \ee

If we have $m$ external soft gravitons then the leading soft term will have $m$ powers
of soft momentum in the denominator. For this each external soft particle must attach
to a nearly on-shell line.  We again sum over all possible insertions, including multiple
insertions on a single external line in all possible order. This leads to a 
generalization of \refb{etwosoftfieldtree} of the form:
\ben \label{esoftgenfieldtree}
\Gamma^{(N+m)}(\varepsilon^{(1)},  k_1, 
\ldots , \varepsilon^{(m)}, k_m; p_1, \ldots , p_N)
&=&\prod_{\alpha=1}^m 
\left[\sum_{i=1}^N {1\over p_i\cdot k_\alpha}  \, \varepsilon^{(\alpha)}_{\mu\nu} 
\, p_i^\mu \, p_i^\nu\right] \, 
\Gamma^{(N)}(p_1, \ldots , p_N)
\nonumber \\ &&
+ \hbox{less singular terms}\, .
\een

\sectiono{Generalizations} \label{sgen}

In this section we shall discuss possible generalizations of our result. One
immediate generalization is the derivation of the leading 
soft photon theorem for superstring tree amplitudes 
for arbitrary number of external states\cite{low}. 
The analysis is very similar
to that in \S\ref{smulti}.
We determine the leading order coupling of a soft photon
of polarization $\ve_\mu$ by replacing the momentum  
$q_\mu$ by $q_\mu - Q\, \ve_\mu$ 
in the expression for the kinetic term. Here
$Q$ denotes the
charge carried by the particle. This gives the analog of \refb{essone1}:
\be \label{essone1Q}
2\, {i} \, Q\, \ve^{\mu} p_{i\, \mu}  \prod_{j=1}^{n_i}\eta^{\mu_j \nu_j}\, .
\ee
Using this we can follow the same procedure as given in \S\ref{stree},
\S\ref{smulti} to derive 
the multiple leading soft photon theorem for $m$ soft photons  of polarizations
$\ve^{(1)},\ldots , \ve^{(m)}$ and momenta $k_1,\ldots , k_m$ attached to an
amplitude with $N$ finite energy particles:
\ben \label{esoftgenfieldcharge}
\Gamma^{(N+m)}(\varepsilon^{(1)},  k_1, 
\ldots , \varepsilon^{(m)}, k_m; p_1, \ldots , p_N)
&=&\prod_{\alpha=1}^m 
\left[\sum_{i=1}^N {1\over  p_i\cdot k_\alpha}  \, Q_i\, \varepsilon^{(\alpha)}_{\mu} 
\, p_i^\mu \right] \, 
\Gamma^{(N)}(p_1, \ldots , p_N)
\nonumber \\ &&
+ \hbox{less singular terms}\, .
\een
For soft photons, we do not expect any universal result beyond the leading term 
since soft photons can couple to the rest of the fields via non-minimal
coupling involving the field strength, and these interactions cost only one
power of soft momentum. Therefore \refb{esoftgenfieldcharge} gives the most
general universal soft photon theorem.

A more interesting generalization would be to extend the analysis of \S\ref{stree}
to derive the subleading soft graviton theorem to all orders in string perturbation
theory for all finite energy external states.  
For this let us restrict the discussion to the cases where the number of
non-compact space-time dimensions is five or more so that the amplitudes are
free from infrared divergences.\footnote{In four space-time dimensions the 
subleading soft graviton theorem is known to be corrected due to 
infrared divergences\cite{1405.1015}.}
In this case the contribution to the amplitude with
$N$ finite energy external particles and one soft graviton
will still 
be given by the sum of the contributions shown in Fig.~\ref{f1fieldtree} and
Fig.~\ref{f3fieldtree}, but in Fig.~\ref{f1fieldtree} the three point vertex 
describing the coupling
of the soft graviton to the finite energy particle is now
the full 1PI vertex and the internal
propagator carrying momentum $p_i+k$ is now the full (finitely) renormalized 
propagator, and in Fig.~\ref{f3fieldtree} the blob labelled
$\wt\Gamma$ now denotes the full Green's
function with external propagators removed and the contributions of the form given
in Fig.~\ref{f1fieldtree} subtracted.
The subleading 
contribution coming from Fig.~\ref{f3fieldtree} will continue to have 
the same form as given in
\refb{esecondtree}.\footnote{In this context note that even though 
we have used a notation in which the
tensor fields carry space-time indices, we could have also used a notation in which they
carry flat tangent space indices by multiplying the tensors by
the symmetric square root of the
inverse metric for each index and treating these as independent field variables. 
This would make some of the intermediate steps in the
analysis different, {\it e.g.} in \refb{esecondtree} the second term will be absent
and in \refb{ecov}  we would have to use the spin connection instead of Christoffel
symbol for defining covariant derivatives. However
the final result would remain unchanged and this formalism would also
be suitable for application to states carrying spinor indices, at least for the evaluation
of the contribution from Fig.~\ref{f3fieldtree}.}
The main difficulty is in the determination of the contribution
to Fig.~\ref{f1fieldtree}  to the first subleading order, since the renormalized two point
function does not have the simple form given in \refb{eactionsft} and 
consequently the 
coupling of the soft graviton to the finite energy particles 
also have a more complicated form.
A promising avenue will be to try to show that the 
quadratic term in the fields in the 1PI 
effective action\cite{1703.06410} 
of superstring field theory
can be brought to the diagonal form given in \refb{eactionsft} after a
field redefinition. In that case we
can use this 1PI effective action to derive the subleading soft graviton theorem 
following the analysis of \S\ref{stree}. This should certainly be possible for scalar
fields, but it is not clear if this can also be achieved for general tensor fields. 
Another possibility -- which should also apply to Ramond sector external states --
will be to work with the renormalized kinetic term as given in \cite{1703.06410}, but use the
gauge invariance of the 1PI effective action to show that the general results of
\S\ref{stree} still hold. 

In this context it is useful to note that 
the contribution to Fig.~\ref{f3fieldtree} given in
\refb{esecondtree} takes the form of a differential operator acting on the amplitude
without the soft particle, and that the differential operator is `local' in the sense that it
contains a sum of terms each of which involves the soft momentum and the momentum 
of one finite energy external state. The contributions from Fig.~\ref{f1fieldtree} also
will have this form. These were the key ingredients based on which
\cite{1406.6574} determined the form of the subleading 
soft graviton theorem up to some overall
normalization constants multiplying the terms that act non-trivially on the polarization
tensors of the external particles. 
The new feature here is the possibility of mixing between states carrying different tensor
indices at the same mass level, including mixing between physical and unphysical states.
Therefore it is conceivable that by gaining some basic knowledge of the contribution
to Fig.~\ref{f1fieldtree} involving the soft graviton coupling to the external state and
the renormalized propagator
one will be able to prove the subleading soft graviton theorem for these amplitudes.
Such arguments should be applicable to all external finite energy states, both in the  NS and
R-sector.

\bigskip

\noindent {\bf Acknowledgement:}
This work was
supported in part by the 
DAE project 12-R\&D-HRI-5.02-0303 and J. C. Bose fellowship of 
the Department of Science and Technology, India.

\end{document}